\documentclass[floatfix, twocolumn]{aastex62}

\usepackage{longtable}
\usepackage{graphicx}
\usepackage{amsmath,amssymb}
\usepackage{color}
\usepackage{units}
\usepackage{epstopdf}
\usepackage{hyperref}
\usepackage{multirow}
\usepackage{chngcntr}

\usepackage{subfigure}
\usepackage{booktabs}

\newcommand{\beq}{\begin{equation}}
\newcommand{\eeq}{\end{equation}}
\newcommand{\bdm}{\begin{displaymath}}
\newcommand{\edm}{\end{displaymath}}

\definecolor{Gray}{gray}{0.9}
\definecolor{orange}{rgb}{0.9,0.5,0}

\providecommand{\keywords}[1]
{
  \small	
  \textbf{\textit{Keywords---}} #1
}

\graphicspath{{./plots/}}

\begin{document}

\title{A luminosity distribution for kilonovae based on short gamma-ray burst afterglows}

\author{Stefano Ascenzi}
\affil{INAF Osservatorio Astronomico di Roma, Rome, Italy}
\affil{Dip. di Fisica, Universit\'a di Roma Sapienza, P.le A. Moro, 2, I-00185 Rome, Italy}
\affil{  Universit\`a di Roma Tor Vergata, Via della Ricerca Scientifica 1,
  I-00133 Roma, Italy}
 \affil{DARK, Niels Bohr Institute, University of Copenhagen, Blegdamsvej 17, 2100, Copenhagen, Denmark}
 \affil{Gran Sasso Science Institute, Viale F. Crispi 7, L'Aquila, Italy}
\author{Michael W. Coughlin}
\affil{Division of Physics, Math, and Astronomy, California Institute of Technology, Pasadena, CA 91125, USA}
\author{Tim Dietrich}
\affil{Nikhef, Science Park 105, 1098 XG Amsterdam, The Netherlands}
\author{Ryan~J.~Foley}
\affil{Department of Astronomy and Astrophysics, University of California, Santa Cruz, CA 95064, USA}
\author{Enrico Ramirez-Ruiz}
\affil{Department of Astronomy and Astrophysics, University of California, Santa Cruz, CA 95064, USA}
\affil{DARK, Niels Bohr Institute, University of Copenhagen, Blegdamsvej 17, 2100, Copenhagen, Denmark}
\author{Silvia Piranomonte}
\affil{INAF Osservatorio Astronomico di Roma, Rome, Italy}
\author{Brenna Mockler}
\affil{Department of Astronomy and Astrophysics, University of California, Santa Cruz, CA 95064, USA}
\affil{DARK, Niels Bohr Institute, University of Copenhagen, Blegdamsvej 17, 2100, Copenhagen, Denmark}
\author{Ariadna Murguia-Berthier}
\affil{Department of Astronomy and Astrophysics, University of California, Santa Cruz, CA 95064, USA}
\affil{DARK, Niels Bohr Institute, University of Copenhagen, Blegdamsvej 17, 2100, Copenhagen, Denmark}
\author{Chris L. Fryer}
\affil{Center for Theoretical Astrophysics, Los Alamos National Laboratory, Los Alamos, NM 87545, USA}
\author{Nicole M. Lloyd-Ronning}
\affil{Center for Theoretical Astrophysics, Los Alamos National Laboratory, Los Alamos, NM 87545, USA}
\affil{University of New Mexico, Los Alamos; Los Alamos, NM 87544 US}
\author{Stephan Rosswog}
\affil{The Oskar Klein Centre, Department of Astronomy, AlbaNova, Stockholm University, SE-106 91 Stockholm, Sweden}

\begin{abstract}
The combined detection of a gravitational-wave signal, kilonova, and short gamma-ray 
burst (sGRB) from GW170817 marked a scientific breakthrough in the field of multi-messenger astronomy. 
But even before GW170817, there have been a number of sGRBs with possible associated kilonova detections. 
In this work, we re-examine these ``historical'' sGRB 
afterglows with a combination of state-of-the-art afterglow and kilonova models. 
This allows us to include optical/near-infrared synchrotron emission produced by the sGRB as well as 
ultraviolet/optical/near-infrared emission powered by the radioactive
decay of $r$-process elements (i.e., the kilonova). 
Fitting the lightcurves, we derive the velocity and the mass distribution as well as the composition of the ejected material.
The posteriors on kilonova parameters obtained from the fit were turned into distributions for the peak magnitude of the kilonova emission in different bands and the time at which this peak occurs. From the sGRB with an associated kilonova, we found that the peak magnitude in H bands falls in the range [-16.2, -13.1] ($95\%$ of confidence) and occurs within $0.8-3.6\,\rm days$ after the sGRB prompt emission. In g band instead we obtain a peak magnitude in range [-16.8, -12.3] occurring within the first $18\,\rm hr$ after the sGRB prompt.
From the luminosity distributions of GW170817/AT2017gfo, kilonova candidates GRB130603B, GRB050709 and GRB060614 (with the possible inclusion of GRB150101B, GRB050724A, GRB061201, GRB080905A, GRB150424A, GRB160821B) and the upper limits from all the other sGRBs not associated with any kilonova detection we obtain for the first time a kilonova luminosity distribution in different bands. 

\end{abstract}


\section{Introduction}

Compact binary mergers are the main sources of gravitational waves (GW) in the LIGO-Virgo frequency range and among them binary neutron stars (BNS) and neutron star-black hole (NS-BH) systems play a special role since they are also potential sources of electromagnetic radiation. 
A BNS/NS-BH coalescence in fact could lead to the formation of a BH (or even a NS in BNS case) surrounded by an accretion disk that is expected to power a highly relativistic jet that will produce a short gamma-ray burst (sGRB) lasting few seconds \citep{1989Natur.340..126E, Paczynski1991, 1992ApJ...395L..83N, MocHer1993, LeeRR2007, Nakar2007}. 
The sGRB is then followed by a fading synchrotron cooling afterglow, from the shock of the jet with the external medium. This afterglow is visible in X-rays, optical and radio for days to months after the initial prompt gamma-ray emission \citep{SaPiNa1998}.

Moreover, during the merger, a fraction of the NS matter can be ejected from the system either by tidal torques or hydrodynamical forces. This component of unbound matter, usually called "dynamical ejecta", is highly neutron rich and therefore is a natural site for the synthesis of r-process elements \citep{LaSc1974, LaSc1976},
whose radioactive decay can heat the ejecta and power a thermal ultraviolet/optical/near infrared transient known as \emph{kilonova} (or \emph{macronova}) \citep{ LiPa1998,MeMa2010,RoKa2011,KaMe2017}.
Contrary to the sGRB prompt and afterglow emission this transient is expected to be broadly isotropic. This means that in principle after every BNS merger which eject a sufficent amount of matter and every NS-BH mergers leading to the NS disruption\footnote{During a NS-BH merger the NS disruption is not guaranteed. Whether it happens or not depends on the dense matter EOS and on the system's parameters, such as the masses of the compact objects and the BH's spin. In general low NS compactness, low BH masses and high spins favour the NS disruption \citep{PannaraleOhme2014}.},
we could expect to observe a kilonova regardless of the orientation of the system \citep{RoKa2011}.

This characteristic, along with a peak in the bolometric lightcurve of $10^{40}-10^{41}\,\rm erg/s$ at a few hours/days after the merger, makes kilonovae optimal targets for an observational campaigns of GW's electromagnetic counterparts \citep{MeBe2012}.
The observational features of kilonovae depend mainly on the 
mass, velocity, and composition of the ejecta. These parameters are in turn correlated with the equation of state (EOS) of neutron stars (NS) and the mass ratio of the binary
\citep{BaBa2013,PiNa2013,AbEA2017b,BaJu2017,DiUj2017,RaPe2018}. A further crucial parameter is the matter opacity, which strongly influences the spectral range of the emission, the peak luminosity and the time at which the peak occurs \citep{GrKo14}. The matter opacity depends on the fraction of lanthanides (produced in r-process nucleosynthesis) within the ejecta, since the bound-bound opacity of these elements dominates all the other contributions.
Dynamical ejecta may also consist of more than one component of matter characterized by different lanthanide fractions and thus different opacities. The lanthanide free ejecta would generate a bluer and faster evolving transient known as \emph{blue kilonova} \citep{MeFe2014,PeRo2014} while the lanthanide rich ejecta would be responsible for the classical \emph{red kilonova} \citep{KaBa2013}. These multiple components arise from different ejection mechanisms: the matter ejected by tidal torques, being particularly neutron rich, is expected to be rich in lanthanides, while that expelled by hydrodynamical forces (\emph{i.e.} the matter squeezed in the contact interface between the two NS or driven by turbulent viscosity \citep{Radice:2018ghv}) would be lanthanides free, since the increase of temperature due to shock-heating reflects in changing the $\beta$-equilibrium in favor of a less neutron rich mixture \citep{WaSe2014, Rosswog2015}.

A further contribution to the kilonova may come from matter expelled from the accretion disk through winds driven by neutrino energy, magnetic fields, 
viscous evolution and/or nuclear recombination energy \citep{FrWo1999,DiPe2002,LeeRRPa2005,MePi2008,DeOt2008,LeRa2009, FeMe2013,PeRo2014,SiRi2014,JuBa2015,CiSi2015}. 
This component of matter is expelled after the dynamical ejecta and it is expected to travel with lower velocity ($\sim 0.05\,c$ against $0.1-0.3\,c$ of dynamical ejecta). Its lanthanide fraction decreases with increasing neutrino irradiation from the disk and the merger remnant, which is high if the remnant is a fast spinning BH and is maximum if the remnant is a long lived NS \citep{KaFe2015}.

The relative contribution of each component depends on the source properties including the binary mass ratio and the nuclear equation of state 
\citep{RoLi1999,BaBa2013,HoKi13,LeLi2016,RaGa2016,DiUj2017,SiMe2017,AbEA2017f}.

All of these three electromagnetic components (sGRB, afterglow, and kilonova) described above have been observed \citep{LVC2017, Goldstein2017, SavFer2017, Shapee2017,Coulter2017,KiFo2017, DroutPiro2017, MuRR2017, PiDa2017, TaLe2017, SmCh2017, ChBe2017, TrPi2017, MaBe2017, HaNy2017, HaCo2017, DavanzoCampana2018, GhirlandaSalafia2018} following GW170817, the first BNS merger event observed by the LIGO Scientific \& Virgo  Collaborations on the 17th August 2017 \citep{AbEA2017b}. 
The kilonova associated with GW170817 
(named AT2017gfo) showed a peak in the bolometric luminosity of $\sim \rm{few}\, 10^{41} \rm erg/s$ in the first $36\, \rm hr$ after the merger and a very rapid spectral evolution from blue to red \citep{TaLe2017,PiDa2017,SmCh2017}. Although this event is the first unambiguous detection of a kilonova, a few candidates, appearing as near-infrared excesses emerging late time from sGRB afterglow lightcurves, have been identified in the recent past. The first to be discovered and probably the most interesting of them was found in association with GRB130602B \citep{TaLe2013,BeWo2013}. Subsequently other two candidates have been identified in association with GRB050709 \citep{YaJi2015} and GRB060614 \citep{JinLi2015}. Contrary to the case of GW170817/AT2017gfo these claimed detections consist on a single photometric point and the lack of any spectrum makes it impossible to clearly assess the nature of these excesses.  In addition, the concurrent X-ray excess in some of these events, e.g. GRB080503,
GRB130603B, suggest that the near-infrared excess could be explained by shock heating and not kilonova emission \citep{kasliwal17}.  Nevertheless, their chromatic nature along with the time and luminosity at which they have been observed makes the kilonova interpretation plausible.

In this article, we are interested in 
measuring the relative contributions of the afterglow and the kilonova. 
The kilonova is distinguishable with its nearly isotropic emission, bolometric luminosity, and color evolution \citep{MeMa2010,RoKa2011,KaBa2013,BaKa2013,TaHo2013,KaFe2015,BaKa2016,Me2017}.
For our analysis, we combine state-of-the-art afterglow and kilonova models and fit them to optical/near-infrared (NIR) short GRB data. We use the optical/NIR data to understand the spectral parameters of the afterglow and determine whether there is any excess light from a kilonova. We use a parameterized surrogate model presented in \citet{CoDiDo2018} and based on simulations from \cite{KaMe2017} of AT2017gfo for the kilonova and a structured-jet model for short GRBs.
We use then the obtained distributions to produce for the first time a kilonova luminosity distribution in different filters. We also calculate for each kilonova event the contribution of r-process element local density with an analysis similar to that performed by \citet{AbEA2017f} and compare the results with the Solar system measures. 

The paper is organized as follows: in Sec. \ref{sec:DataTechnique} we describe our data sample and kilonova and SGRB models employed in the data fitting. In Sec. \ref{sec:results}
we present the results of our analysis, which comprise the distribution of mass, velocity and lanthanides fraction for all the kilonovae events, the peak luminosity distribution for all kilonovae events as well as the upper limits placed by the kilonovae non-detections, the luminosity distribution of kilonovae in different filters and the contribution to the local r-process elements density for each event. Finally in Sec. \ref{sec:summary} we briefly summarize our analysis and draw the conclusion of our work. 

\section{Data and analysis technique}
\label{sec:DataTechnique}
We begin describing our sGRB sample. This comprises the events GRB130603B
\citep{TaLe2013,BeWo2013}, GRB140903A \citep{TrSa2016},
GRB060614 \citep{ZhZh2007,JinLi2015,YaJi2015}, GRB050709 \citep{FoFr2005,Hjorth2005,Covino2006,Jin2016} , GRB061201 \citep{Stratta2007}, GRB050724A \citep{BePr2005,MalesaniCovino2007}, GRB150101B \citep{FoMa2016, TrRy2018}, GRB080905A \citep{NicuesaGuelbenzu2012,RoWi2010}, GRB070724A \citep{BeCeFo2009,Kocevski2010}, GRB160821B \citep{KaKoLau2017,JinWang2018}, and GRB150424A \citep{TanLevFru2015, JinWang2018}. 
In addition to the short GRBs here, we include measurements from GW170817 (GRB170817A) \citep{AbEA2017e}. 
This sample is a subset of the  \citet{GoLe2018} sample, which includes all SGRBs with 
measured redshift $z \le 0.5$ and from which we selected only the events with an optical/NIR afterglow detected  (not just upper limits). This cut in reshift is motivated by the fact that for $z > 0.5$ the faint kilonova emission would be unlikely detected by present and upcoming telescope facilities. Nevertheless, this limit is much deeper than the LIGO-Virgo horizon at design sensitivity for BNS and NS-BH mergers \citep{AbEA2010}. We excluded, as \citet{GoLe2018} did, also GRB061006, GRB071227 and GRB170428 due to their too luminous host galaxies.
All the photometric data have been corrected for the Milky Way extinction. 
In Table \ref{tab:all_grbs}  the salient information of all the GRBs (and GW170817) in the sample have been summarized. For the cases with a kilonova detection (or claimed detection) the ejecta mass and lanthanide fraction inferred from our analysis are also furnished.

As described above, sGRB afterglows are typically modeled as a decelerating and decollimating relativistic jets producing synchrotron emission. 
From numerical simulations and the analyses of GW170817, 
slow-moving cocoon \citep{NaHo2014,LaLo2017,KaNa2017,MoNa2017, MuRR2017}
and Gaussian structured
jet \citep{TrPi2017} models seem to be preferred, while a universal jet structure 
seems to be disfavored~\citep{TrPi2017,KaNa2017}. 
In the Gaussian structured jet case, energy drops as $E(\theta)=E_0\exp[-\theta^2/(2\theta^2_c)]$ up to a truncating angle $\theta_w$, where $E_0$ is the isotropic equivalent energy and $\theta_c$ is the opening angle. In the following, we will use the formalism adopted by \cite{TrPi2018}, where the Gaussian jet is implemented as a series of concentric top hat jets and the cocoon as a decelerating
shell model which includes ongoing energy injection. 
We use the implementation in \textit{afterglowpy} \citep{RyVa2019}.
This formalism considers also the effect of the viewing angle $\theta_v$, which is thus a further parameter of the model. Concerning the other parameters, we denote as $n$ the number density of the homogeneous environment containing the jet and the power law distribution slope in energy of the electrons undergoing synchrotron emission as $p$. A fraction $\epsilon_E$ contains the post-shock internal energy, while a fraction $\epsilon_B$ contain the shock-generated magnetic field energy.  

For the kilonova model, we use an interpolated surrogate model based on \cite{KaMe2017}, which is described in \citet{CoDiDo2018}. 
The model is parameterized by three variables: the ejecta mass $M_{\rm ej}$, the mass fraction of lanthanides  $X_{\rm lan}$, and the ejecta velocity $v_{\rm ej}$. This model provides a state-of-the-art, parameterized model to test our analysis method.  But it makes a series of assumptions that may ultimately affect our results.  It assumes spherical symmetry and a uniform composition and uses multi-wavelength radiation transport combined with atomic line data to derive the model.  For the isotopes calculated, the atomic data is state-of-the-art.  But, at this time, many of the lanthanide opacities have not been calculated and, like other studies, this model uses a few well-calculated opacities as surrogates for the entire set of lanthanides.  With multiple ejection processes (dynamical ejecta from tidal disruption, winds from an accretion disk and, if the compact object remains a NS, outflows from NS accretion), the ejecta is likely to have a range of compositions and velocity profiles.  In addition, uncertainties in the nuclear physics can produce radioactive isotopes that can significantly alter the radioactive heating, altering the lightcurve \citep{zhu17}.  In addition, this model assumes that all the kilonova energy is furnished by the radioactive decay of the nuclides synthesized during r-process nucleosynthesis and no kind of central engine (\emph{e.g.} magnetar, pulsar, fallback accretion) is taken into account. The inclusion of this further contribution could lead to a widening of the distributions of kilonova parameters and in particular to lower values of ejecta mass as found by \citet{MaIo2018}.  This differences are, to large extent, the cause in the different yield estimates from GW170817 \citep{cote18}. For the analyses that follow, we will show examples where the afterglow and kilonova models are fit separately to the data, as well as examples where we add the models together to create joint distributions.

We compare these models to observational data following \cite{CoDi2017}, i.e., randomized sets of lightcurves are computed for each model, and a $\chi^2$ value is calculated between each model and the data.
For the kilonova model, the priors are taken to be flat between:
$-5 \leq \log_{10} (M_{\rm ej}/M_\odot) \leq 0$, $ 0 \leq v_{\rm ej} \leq 0.3$\,c, 
and $-9 \leq \log_{10} (X_{\rm lan}) \leq -1$.
For the afterglow model, the priors are taken to be flat between:
$0 \leq \theta_v \leq \pi/4$, $0 \leq \theta_c \leq \pi/4$, $0 \leq \theta_w \leq \pi/4$, $49 \leq \log_{10} (E_0/\rm{erg}) \leq 55$, $-4 \leq \log_{10} (n/\rm{g/cm^3}) \leq 0$, $2.1 \leq p \leq 2.5$, $-4 \leq \log_{10} (\epsilon_E) \leq 0$ and $-4 \leq \log_{10} (\epsilon_B) \leq 0$.
The lightcurve fitting code is available at: 
\url{https://github.com/mcoughlin/gwemlightcurves}.

\section{Results}
\label{sec:results}
\subsection{Mass, Lanthanide fraction and Luminosity Distributions}

\begin{figure*}[t]
    \centering
  \includegraphics[width = 0.75\textwidth]{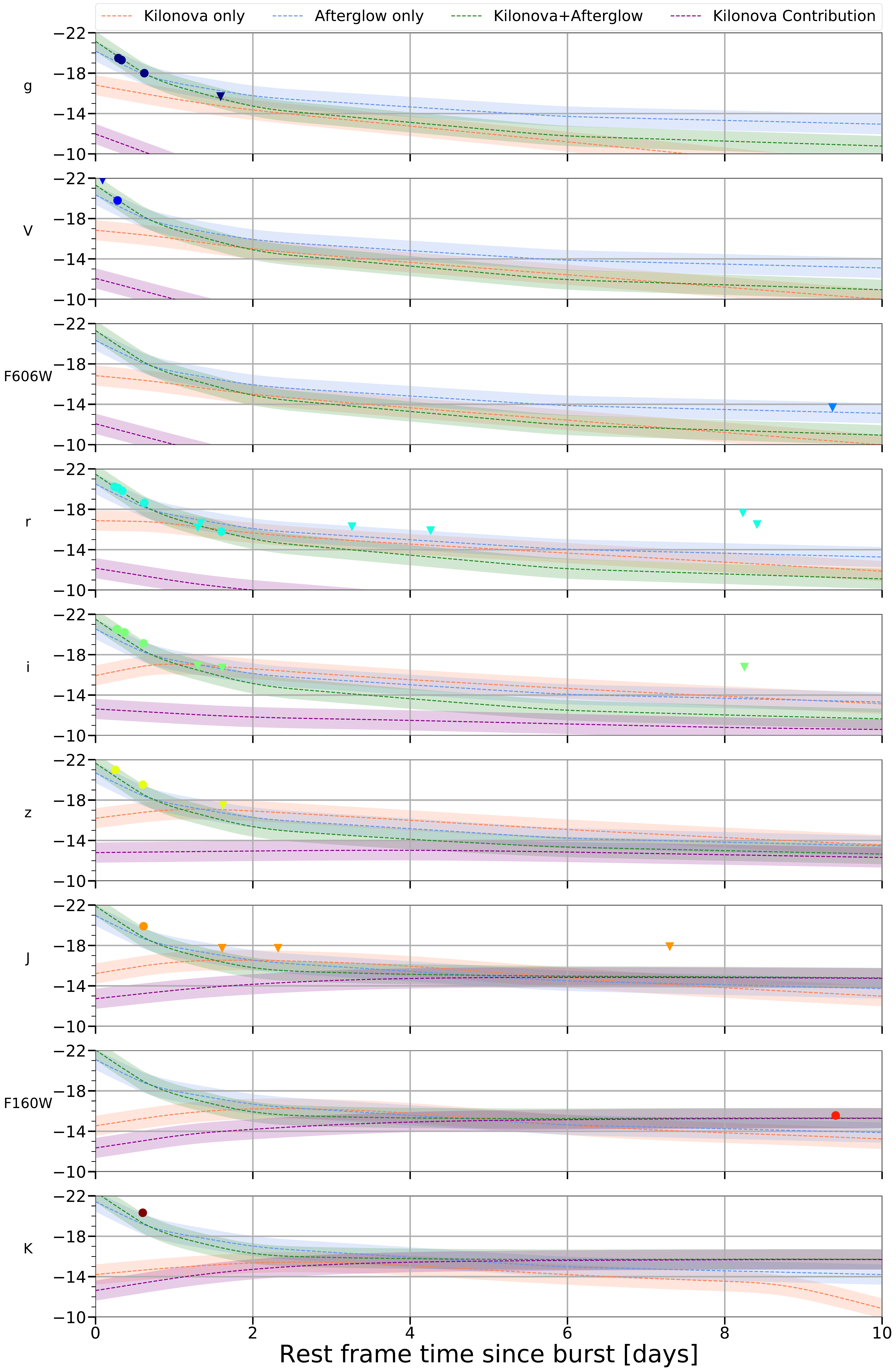}
  \caption{Lightcurves for the Gaussian afterglow model \citep{TrPi2018}, the kilonova model \cite{KaMe2017}, and a sum of the afterglow and kilonova model for GRB130603B. "Kilonova part" denotes the kilonova contribution to the "afterglow+kilonova" model. The shown lightcurves correspond to a maximum likelihood $\chi^2$ fit to the data. All the lightcurves are expressed in AB absolute magnitudes. The circles denote actual detections while the triangles are upper limits.
 }
 \label{fig:GRB130603B}
\end{figure*}

\begin{figure}[t]
\centering
 \includegraphics[width=3.3in]{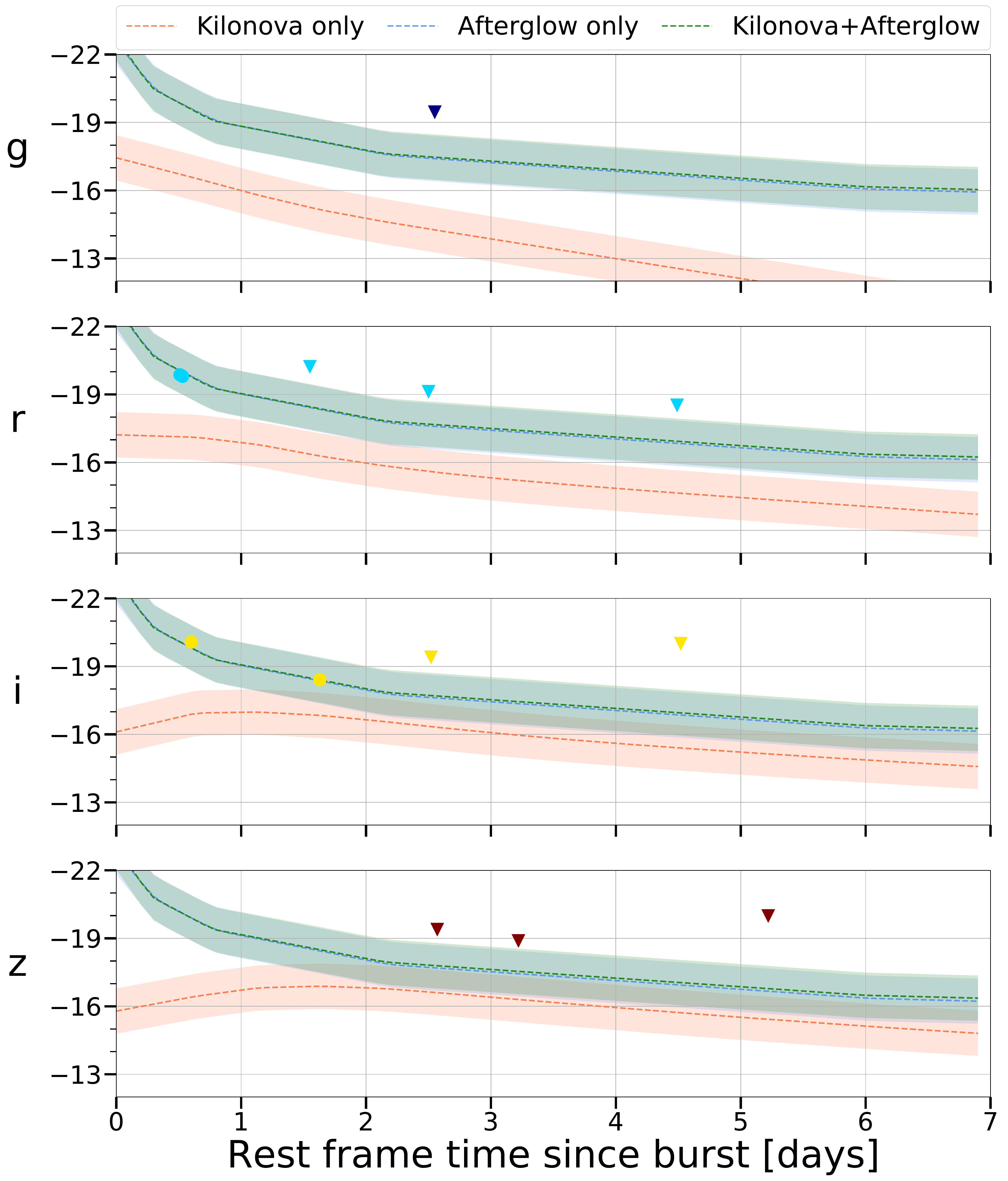}
  \caption{Same as Figure \ref{fig:GRB130603B} for GRB140903A. Afterglow and Kilonova + Afterglow models superimposed exactly and are not distinguishible.}
 \label{fig:GRB140903A_model}
\end{figure}

\begin{figure}[t]
\centering
 \includegraphics[width=3.3in]{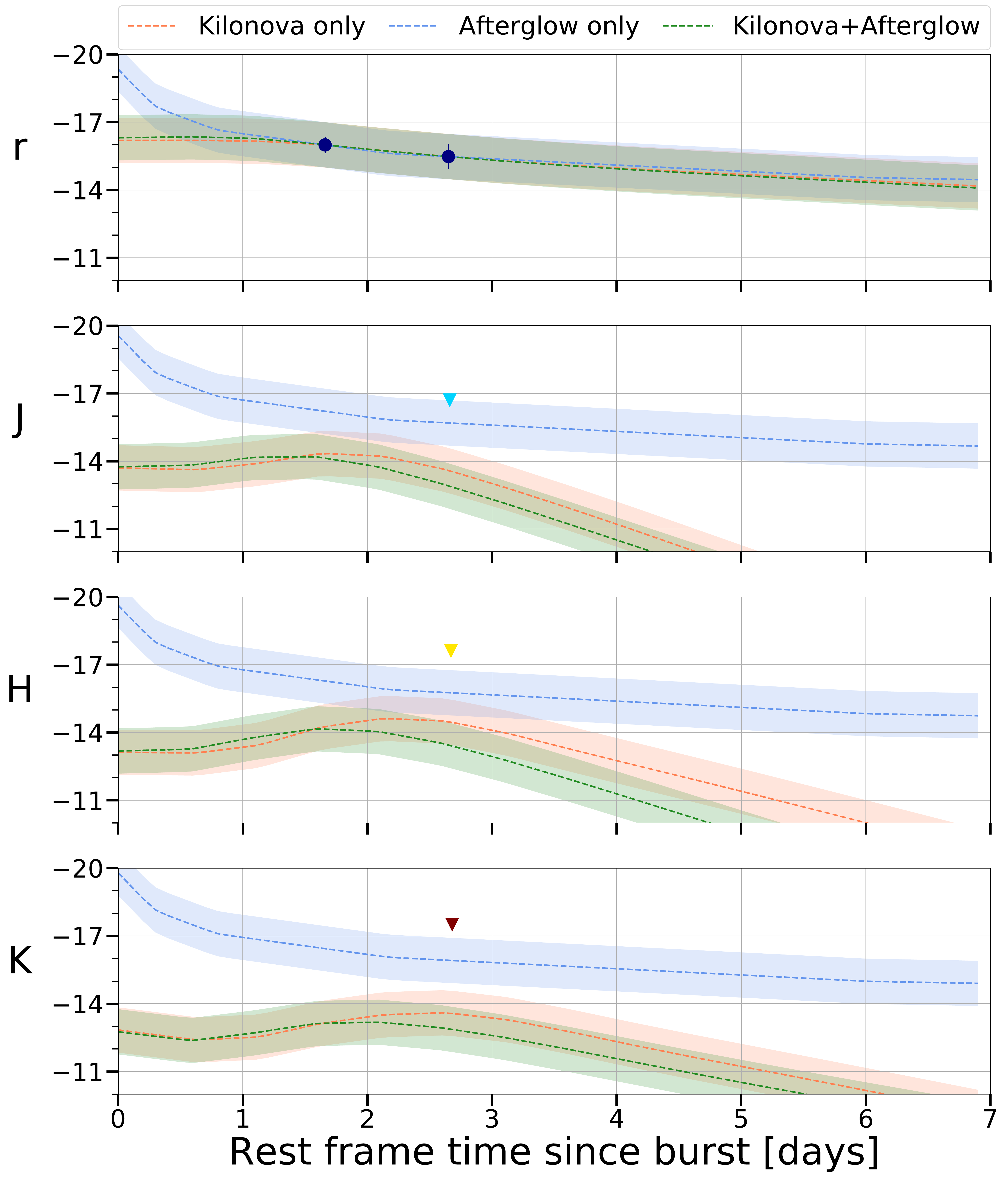}
  \caption{Same as Figure \ref{fig:GRB130603B} for GRB150101B.
  }
 \label{fig:GRB150101B}
\end{figure}

\begin{figure}[t]
\centering
    \includegraphics[width = 0.96\columnwidth]{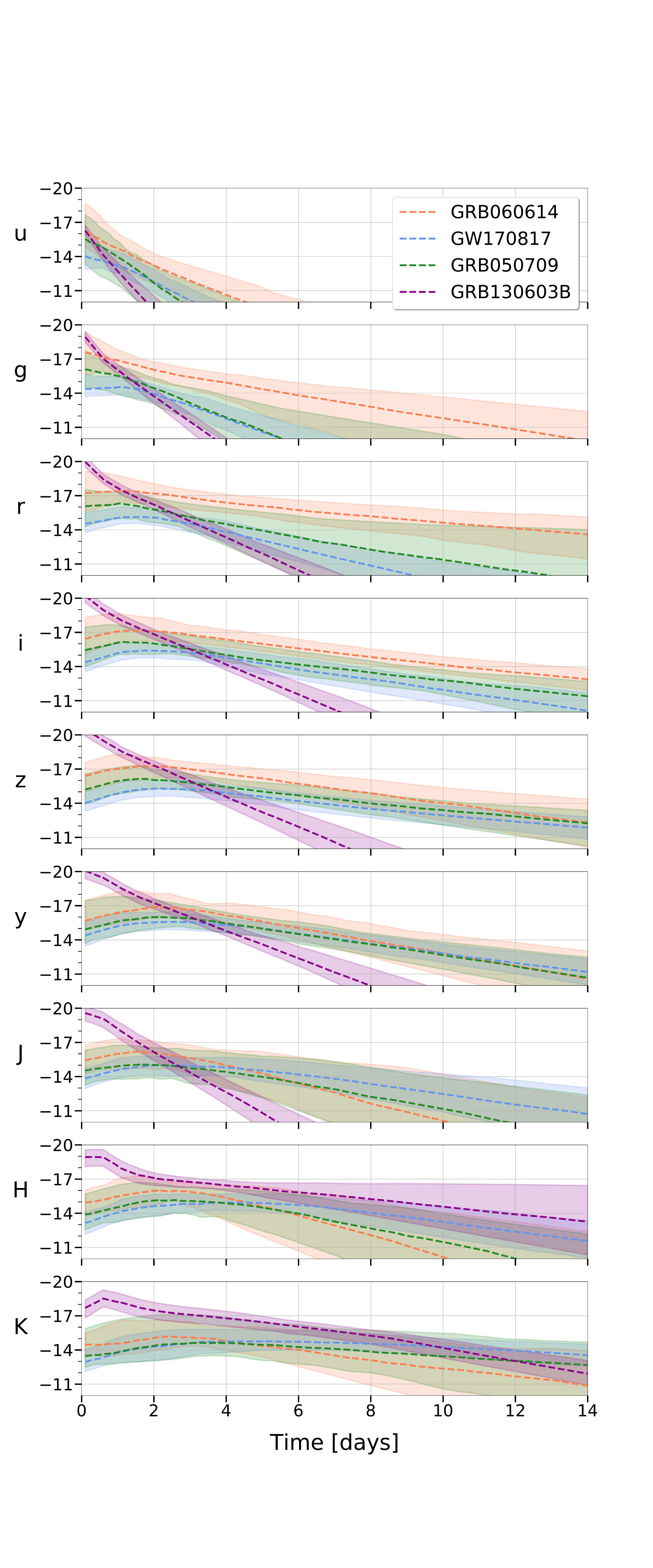}
 \caption{
 Lightcurves obtained by the fitting procedure for the kilonova events.
    The lightcurves are those of the surrogate kilonova model with the exception of GRB130603B where a kilonova+afterglow model have been employed.
 The dashed lines show the median light curve, while the shaded intervals show the 95\% intervals.
    The numbers to the left of the y-axis show the passbands of the observations. 
    }
  \label{fig:fits}
\end{figure}

\begin{figure}[t]
\centering
  \includegraphics[width=\columnwidth]{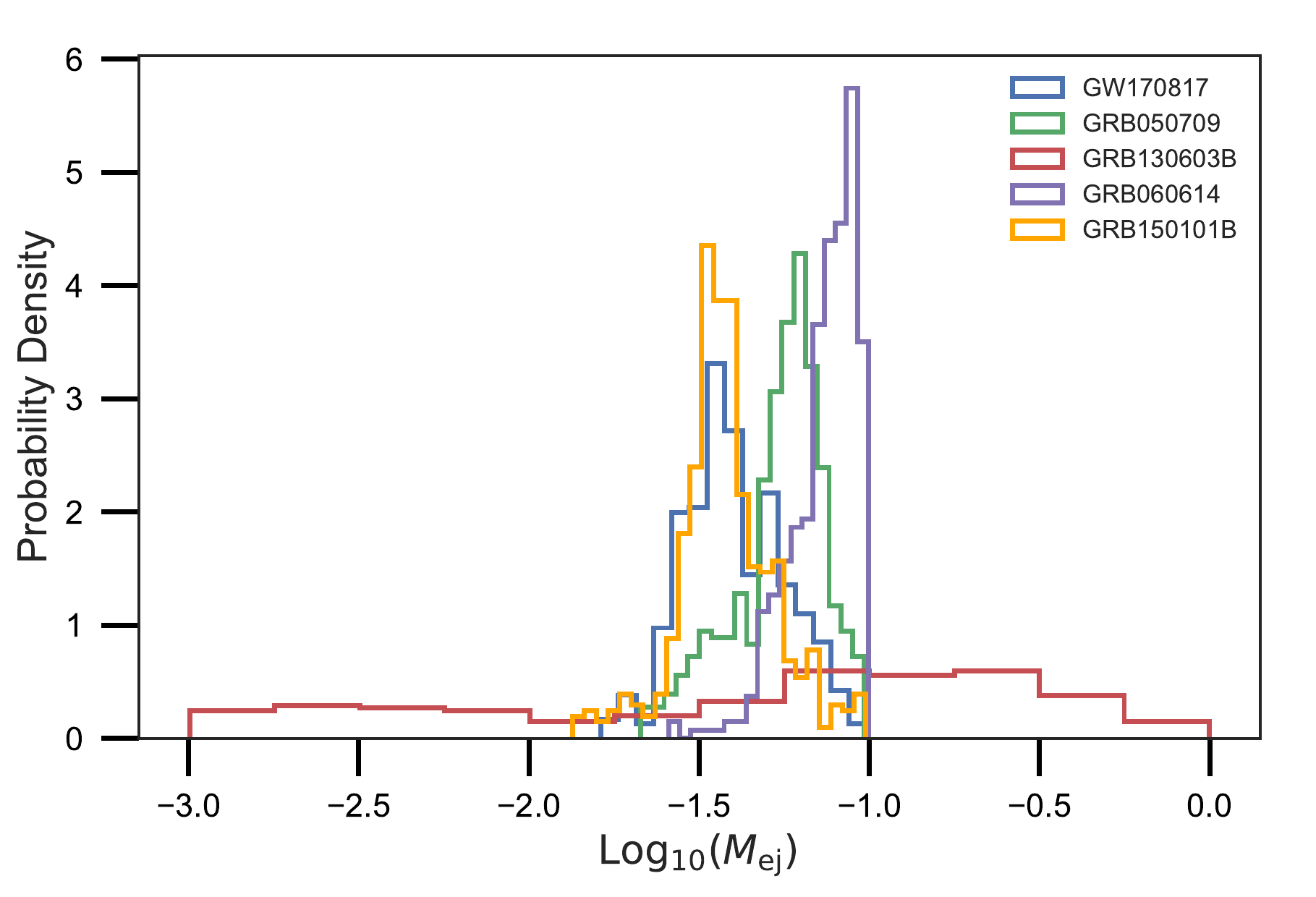}\\
  \vspace{1cm}
  \includegraphics[width=\columnwidth]{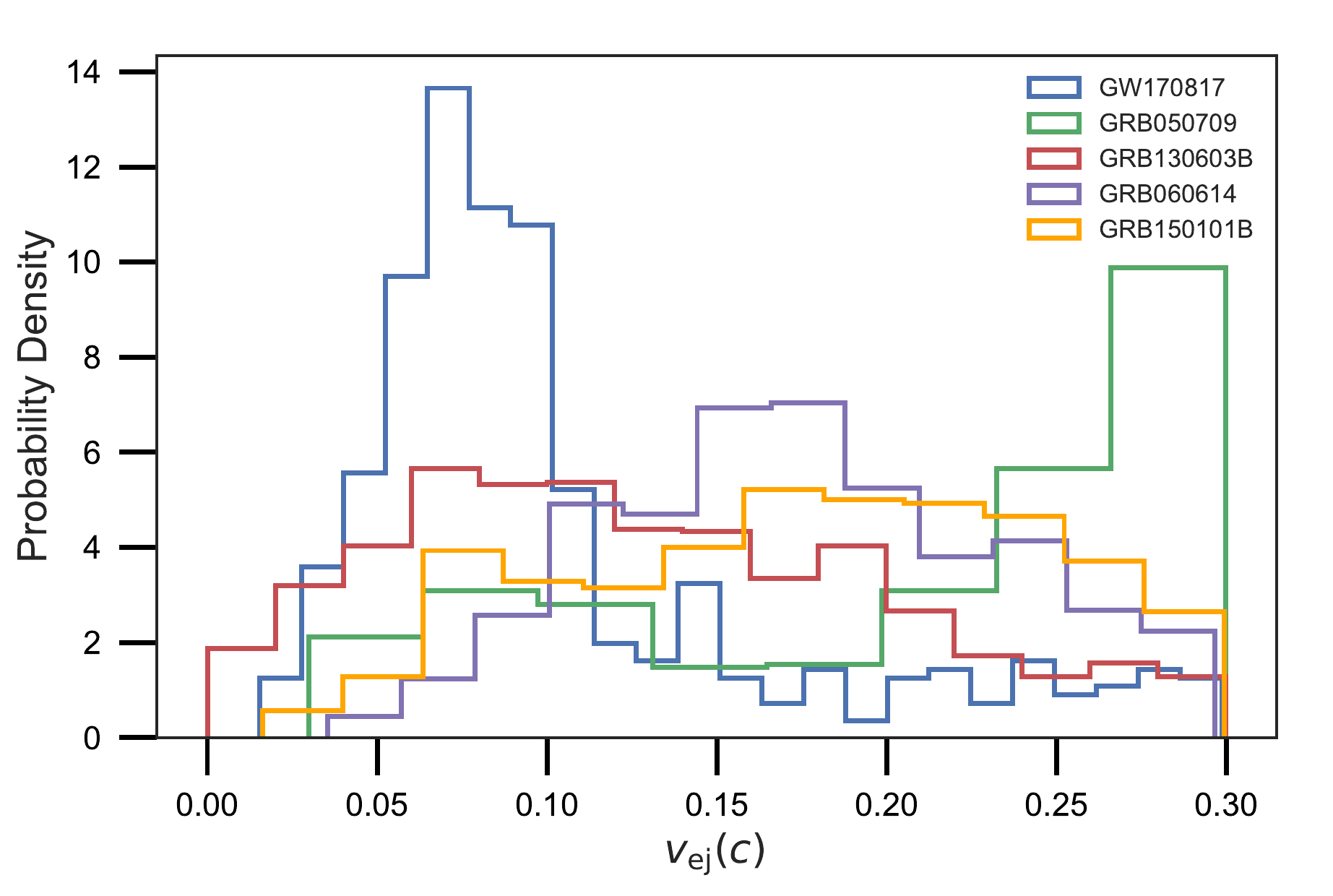}\\
  \vspace{1cm}
  \includegraphics[width=\columnwidth]{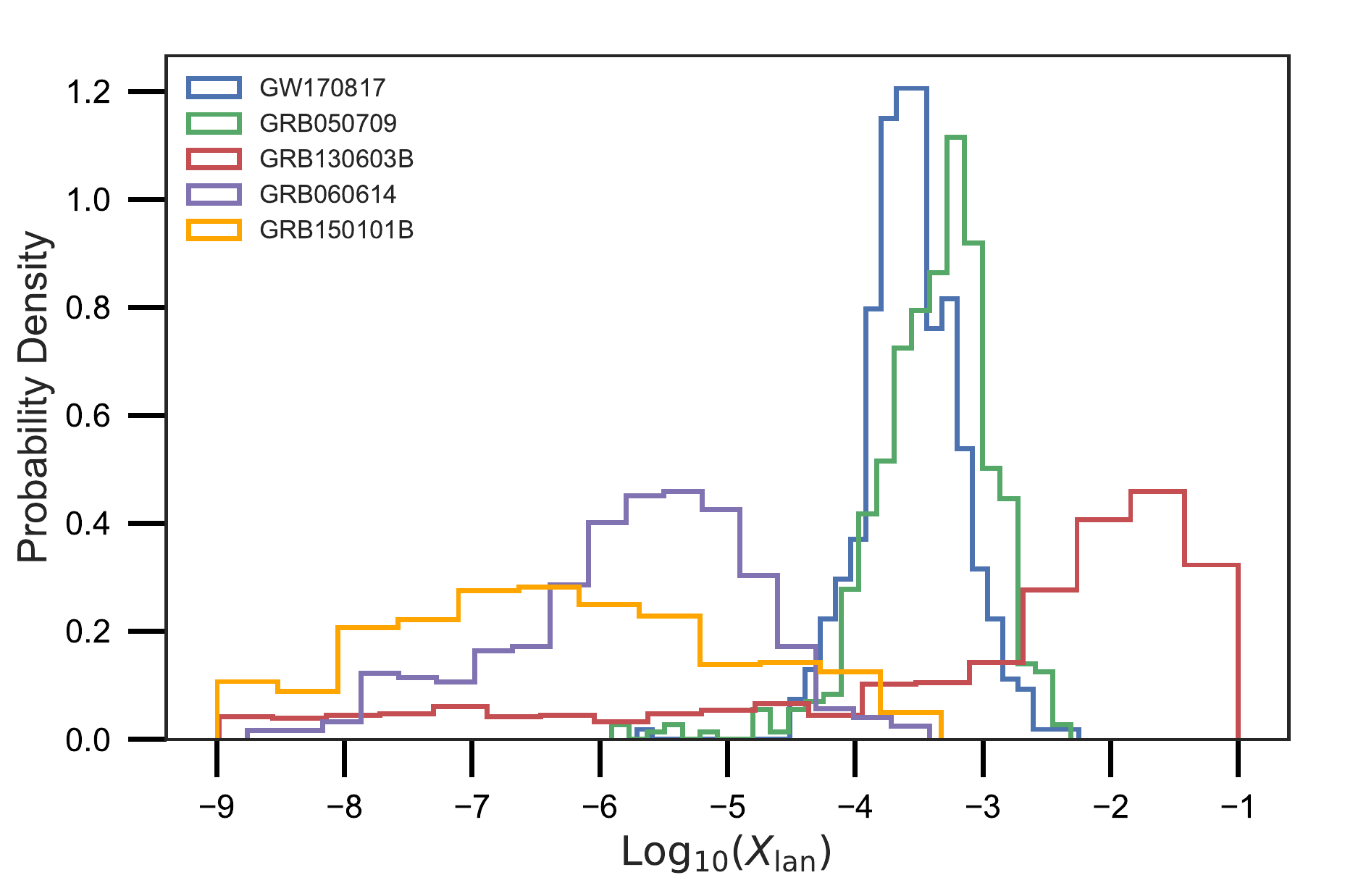}
  \caption{Ejecta mass (top) ejecta velocity (middle) and lanthanide fraction (bottom) estimates based on the GRB sample with an associated kilonova  considered in this paper. 
}
 \label{fig:mass}
\end{figure}

We begin with an analysis of GRB130603B to illustrate the method. 

Figure \ref{fig:GRB130603B} shows the observed data superimposed on different fitted models. Here \emph{afterglow only} denotes the Gaussian afterglow model of \citet{TrPi2018}, \emph{kilonova only} the kilonova model of \citet{KaMe2017} and \emph{kilonova+afterglow} the combination between the two models. Adding a kilonova contribution to the afterglow causes an increase in the optical flux at early times and, as a result, predicts less flux from the afterglow at later times (compared to the afterglow only model).   The additional (relative to the afterglow) contribution necessary from the kilonova to account for the observations is shown by the purple line (denoted as \emph{kilonova contribution}).

In this figure we can see that the kilonova in addition to the afterglow is required to fit the data, as noted at the time of detection \citep{TaLe2013,BeWo2013}. We find that the estimate of the ejecta mass based on the joint analysis is $M_{\rm ej} = 7.46^{+43.97}_{-7.29}\times 10^{-2} M_\odot$.

A different case, shown in Figure \ref{fig:GRB140903A_model}, is that of GRB140903A, where the afterglow fit dominates the performance of the fit. If a kilonova is present here its lightcurve is completely buried in the afterglow lightcurve and any upper limit on ejecta mass would be too high to be informative. 
In fact the fit of the kilonova model results in $M_{\rm ej} \le 7.46 \times10^{-1}\,M_\odot$.

The final scenario is represented by GRB150101B and is shown in Figure \ref{fig:GRB150101B}. In this case the afterglow, the kilonova and the afterglow + kilonova fit perform equally, which means that although we cannot claim a kilonova detection we can put an informative upper limit on the ejecta mass. This measure is equal to $M_{\rm ej} = 3.17^{+3.12}_{-1.56}\times 10^{-2}\, M_\odot$.

We want to use our analysis of the individual short GRBs to make constraints on the luminosity and ejecta mass of the kilonovae.
For some of the short GRBs, such as GRB140903A and GRB050724A, the photometry is such that no (informative) limits on kilonova emission are possible; in other words the analysis gives back the parameters priors.
The ones of most interest to us are the ones which provide some limits like GRB061201 and GRB080905A (as also noted by \citet{GoLe2018}), or even (claimed) detections of kilonovae. These include GRB150101B, GRB050709, GRB130603B, GRB060614 and GW170817/GRB170817A. 

 We first show the lightcurves predicted by the fitting analysis in Figure~\ref{fig:fits}. These lightcurves are similar in concept to those in Gaussian Process Regression, where the lightcurves span the possible extrapolations based on the model, which is in this case the kilonova surrogate model. Only in the case of GRB130603B is the afterglow model added because we need both the afterglow and kilonova components to fit the data. 
 In the plot, the dashed lines show the median lightcurve, while the shaded intervals show the 95\% intervals. 

Figure~\ref{fig:mass} shows the posteriors of the $M_{\rm ej}$, $v_{\rm ej}$ and $X_{\rm lan}$ for the events that we regard as a real kilonova detection, where we included also the recently claimed blue kilonova associated to GRB150101B \citep{TrRy2018}.
They are both broadly consistent in this measurement to what was found for GW170817.
This is not an accident, as there is a significant selection effect in this analysis.
Some afterglows return the priors (see Figure \ref{fig:afterglow_params} for the posterior distributions of the afterglow's parameters), given the significant energies involved; their lightcurves are not informative.
This is the subset with low enough afterglow energies to be consistent with the energies we expect from kilonovae.
Perhaps most interesting that no observations are consistent with measurements lower than $\approx 0.05 M_{\odot}$.
These large masses are  commonly thought  to be less likely to be produced by dynamical ejecta (e.g.~\citealt{BaBa2013,Rosswog2013,HoKi13,DiBe2015}). 
Instead magnetized or neutrino-irradiated wind from a long-lived hypermassive NS remnant prior to its collapse to a black hole is usually invoked \citep{MeTh2018}.  
In general, GRB130603B has the broadest range of possible parameters for a few reasons. As stated previously, GRB130603B is the only one where we include the afterglow model as well. In addition, the main contribution of the kilonova model is to improve the fit to the final data point at about 9.5\,days. For this reason, the posteriors are driven by any kilonova parameters that pass through this set of data points. These are required to achieve a lightcurve sufficiently red to reach a magnitude brighter than $m_{\rm AB}=-16$ and blue such that it is dimmer than $m_{\rm AB}=-14$.
This event is more consistent with large amounts of red ejecta, which could originate from an accretion disk outflow (e.g.~\citealt{MeFe2014,PeRo2014}), just as the blue ejecta.

The distributions of $M_{\rm ej}$, $v_{\rm ej}$ and $X_{\rm lan}$ have been turned into distributions for time of the peak and peak magnitudes in different filters. To this aim, first a distribution of opacities $k$ have been obtained from $X_{\rm lan}$ using a log-linear relation between the two parameter described by the equation: 

\begin{align}
	k(X_{\rm lan}) &=
    \begin{cases}
    	m\log_{10}\bigl(X_{\rm lan}\bigr) + q\quad &X_{\rm lan} \ge 10^{-6}\label{eq:opacity} \\
        k_0 \qquad &X_{\rm lan}\le 10^{-6}
    \end{cases},\\
   \notag \\\notag
    m &= \frac{k_1 - k_0}{\log_{10}\bigl(X_{\rm lan, 1}\bigr)- \log_{10}\bigl(X_{\rm lan, 0}\bigr)},\\
    \notag\\\notag
    q &= \frac{k_0 \log_{10}\bigl(X_{\rm lan, 1}\bigr) - k_1 \log_{10}\bigl(X_{\rm lan, 0}\bigr)}{\log_{10}\bigl(X_{\rm lan, 1}\bigr) - \log_{10}\bigl(X_{\rm lan, 0}\bigr)}, 
\end{align}
where $k_0 = 0.1\,\rm cm^2/g$, $k_1 = 10\, \rm cm^2/g$, $X_{\rm lan, 1} = 10^{-1}$, $X_{\rm lan, 0} = 10^{-6}$. This prescription ensures the opacity to be equal to $k_0$ when $X_{\rm lan}= X_{\rm lan, 0}$ and rise logarithmically to $k_1$ at $X_{\rm lan} = X_{\rm lan, 0}$.

\begin{figure*}[htp]
\centering
 \includegraphics[width=6.5in]{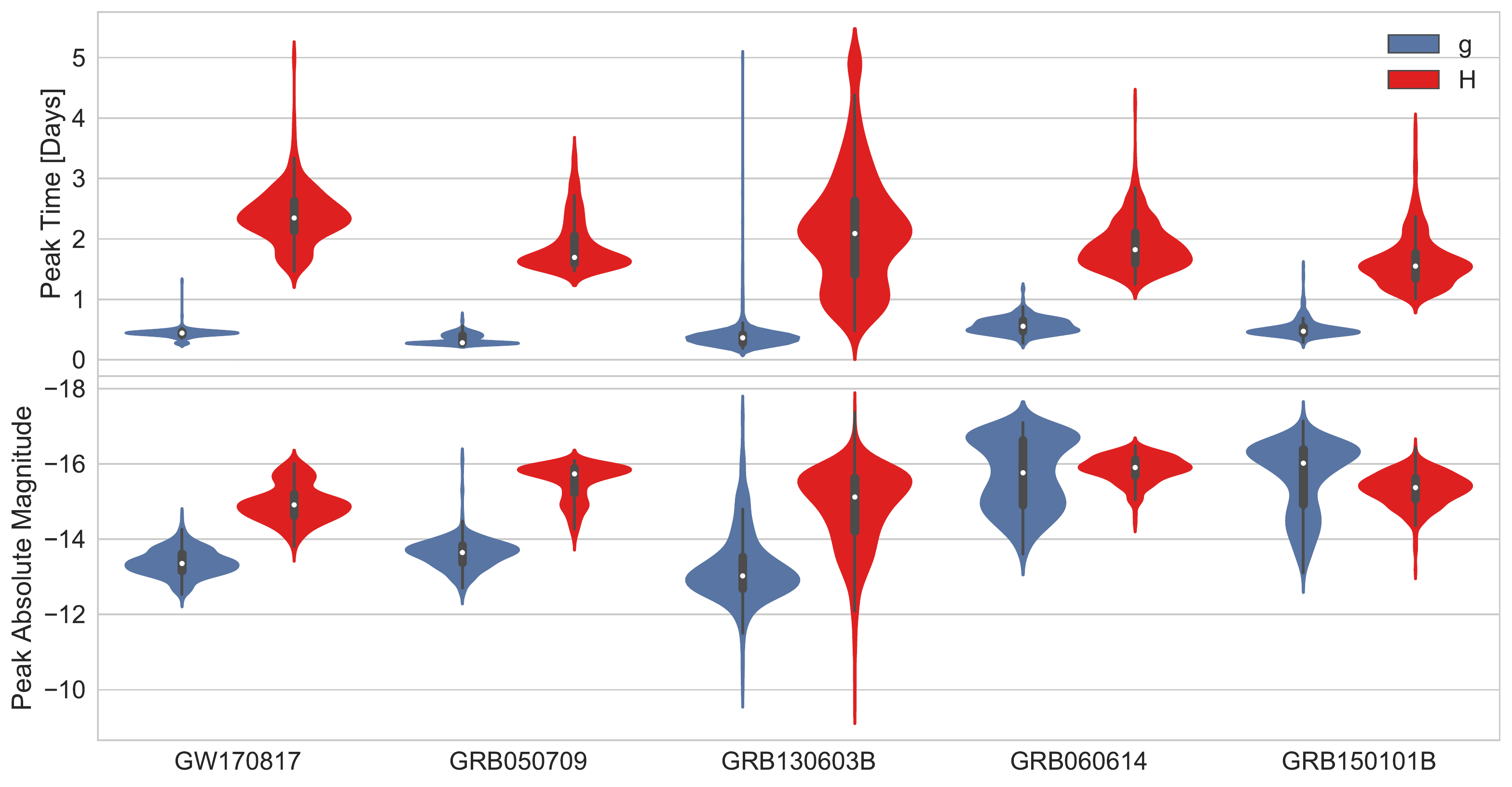}
 \caption{Violin plot showing the peak time (top) and peak magnitude (bottom) for all the events considered in the filters g (blue) and H (red). White dots, 	black bars and black lines represent respectively the median, the interquartile range and the 95\% confidence interval of the distributions. In this plot only the events that lead to the most constraining measures/upper limits on luminosity distribution are shown. For a plot that include the whole sample see Figure \ref{fig:violin_all}.}
 \label{fig:violin}
\end{figure*}

\begin{figure*}[htp]
\centering
	\includegraphics[width=7.0in]{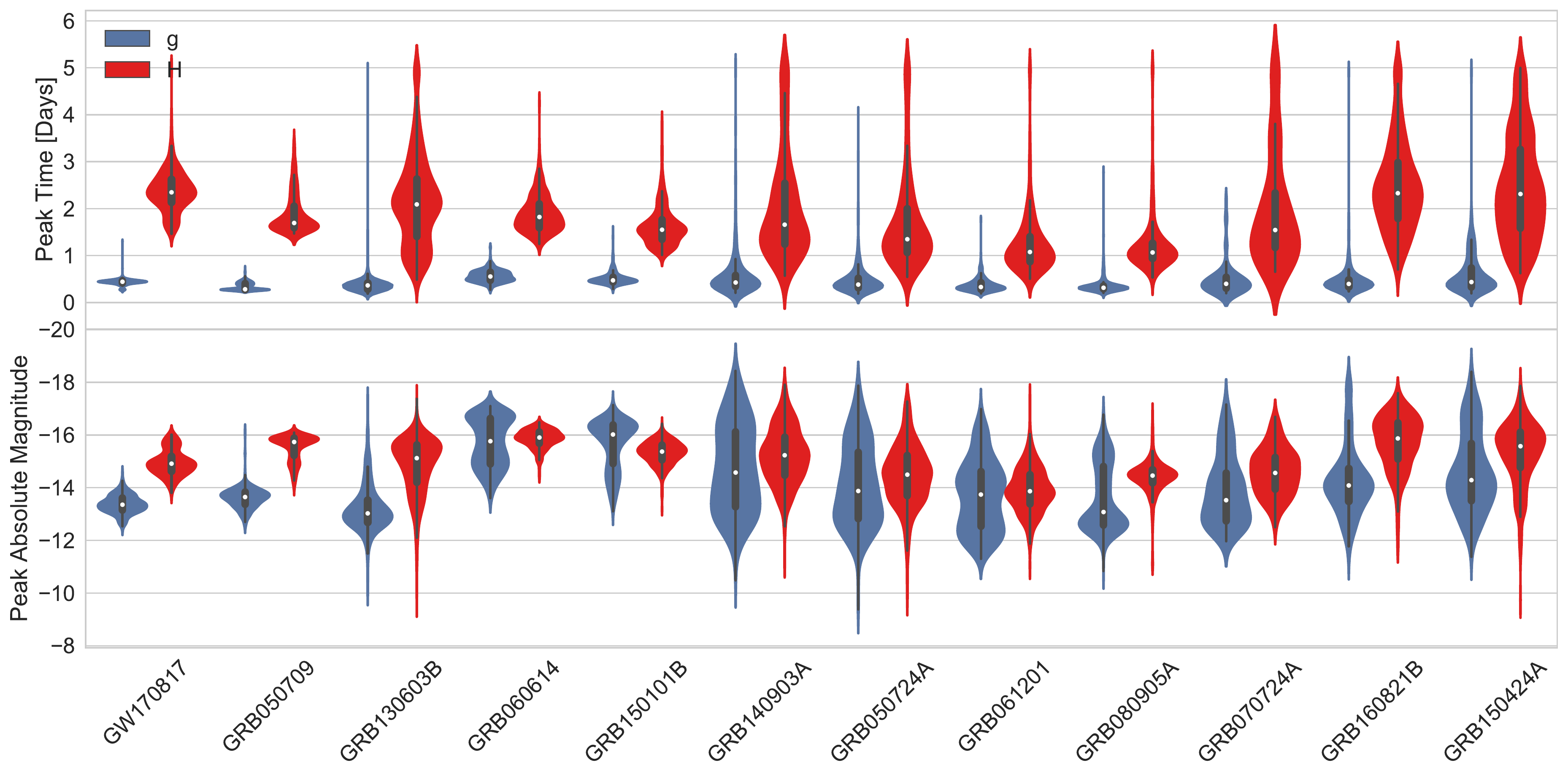}
    \caption{Same as Figure \ref{fig:violin} but with all the events in our sample.}
    \label{fig:violin_all}
\end{figure*}

Then opacities, masses and velocities of the ejecta have been turned into kilonova multicolored lightcurves following the method/model outlined in \citet{Me2017}. Although this is a simple analytical model it reproduces results accurate enough for our analysis. For each lightcurve in different filters the peak of the luminosity, along with the time at which it occurs, has been obtained. We report our results in Figure \ref{fig:violin} and Figure \ref{fig:PeakTimeMag}.

Figure \ref{fig:violin} shows for the most interesting events
(those of Figure \ref{fig:mass}) the distribution of peak time (top panel) and peak absolute AB magnitude (bottom panel) for the filter g (blue) and H (red). The white dot represents the median of the distribution for the given event. The black bars and lines mark respectively the interquartile range and the 95\% of confidence interval of the distribution. 
In Figure \ref{fig:violin_all} we report the same results for the whole sample.

In the Appendix, in Figure \ref{fig:PeakTimeMag}, we present in a 2D peak magnitude-peak time space the probability density distribution (within a 68\% of confidence interval) in the central panel (again for the filter g and H in blue and red respectively) to highlight the correlation between the two parameters. These distributions have been drawn smoothing the discrete data with a Gaussian kernel based density estimation. The top and right panel show the marginalized distributions of peak time and peak magnitude respectively.

Figure \ref{fig:violin} and Figure \ref{fig:PeakTimeMag} show that in all cases the peak time of the emission in g filter lies within few hours after the merger and within the first three days from the merger/GRB prompt emission in the H filter. The H peak magnitude is expected to lie in the range $[-16.2, -13.1]$ (95\% of confidence), while the g filter distribution is broader with a peak magnitude laying in the range $[-12.3, -16.8]$. Events GRB060614 and GRB150101B show a double peaked g luminosity distribution with the smaller peak below the median H luminosity. 
GRB150101B shows also a dominant blue component (referring to the median of the distribution) that results from the low inferred lanthanide fraction. It is worth noticing that recently \citet{TrRy2018} found evidence for a blue kilonova arising from the early time ($t\sim $ 2 days) ultraviolet/optical lightcurve of the GRB150101B afterglow. Although we do not find any firm evidence of a kilonova for this GRB, its contribution is not ruled out and still consistent with the result of our analysis.
 Moreover, the authors found for the kilonova associated with GRB150101B, an ejecta mass $M_{\rm ej}> 0.02\,M_\odot$ and an opacity $k \sim 1\, \rm cm^2/g$ (equivalent to $X_{\rm lan} \sim 10^{-6}$ according to Equation \ref{eq:opacity}) both consistent with our results (see Table \ref{tab:all_grbs}).

Among all the events GRB130603B (the prototype of an afterglow+kilonova fit) shows the largest uncertainties both in peak time and peak magnitude distributions. This follows from the very large inferred distributions of mass and lanthanide fraction (see Figure \ref{fig:mass}), which span the full parameter space. 

If we look at the Figure \ref{fig:PeakTimeMag} we can see that GW170817 and GRB050709 manifest (in the H filter at least) a correlation between peak time and peak magnitude, with the higher luminosities having the lower peaking time. This correlation is absent in the other cases. Again we can interpret this trend in view of Figure \ref{fig:mass}, where it is worth noticing that the two cases considered above are those with the lowest dispersion both in mass and lanthanide fraction. When these parameters are well constrained it is thus natural to expect from the model the event with higher luminosities to peak earlier in time. We expect to observe this kind of trend in future observations with a less limited dataset. 

\subsection{The kilonova luminosity distribution}

We used the distribution of the peak luminosities in Fig. \ref{fig:violin_all}
to build a set of luminosity distributions for each spectral band. We consider first the real kilonovae
events, namely GW170817, GRB130603B, GRB050709 and GRB060614 with the eventual inclusion of GRB150101B. Moreover, recently \citet{Rossi2019} claimed a kilonova association also for GRB050724A, GRB061201, GRB080905A, GRB150424A, GRB160821B, so also them are eventually included\footnote{All these 5 events have been fitted with an afterglow+kilonova model.}. For each of these events from the peak magnitude distribution in the chosen filter we took the median, the 5th and the 95th percentiles of the distribution. In this way for each event the distribution is reduced to three values representing the median, the upper and lower limits on kilonova peak magnitude. 
Now we turn to all the other events that are not associated to any kilonova. 
For these cases we take only the 5th percentile of the distribution as the peak magnitude upper limit and we set the median and the lower limit to an infinity magnitude (corresponding to 0  luminosity). In this way for the whole sample we obtain three sets of values that we use to draw three different cumulative distribution functions representing the kilonova luminosity distribution in three limiting cases: the optimistic case, obtained from the upper limits, the pessimistic case obtained from the lower limits and the median case obtain by the median of the distributions. Whether our luminosity distribution represents also a proper luminosity function of kilonovae depends on the selection effects of our sample. The selection effect in our case is represented by the detection of an afterglow associated to a kilonova. In this way our luminosity distribution would correspond to the luminosity function for the kilonovae associated with an observed afterglow (which means on-axis orientation and systems with non-choked jets). It is worth noticing that the luminosity function defined in this way is the cumulative distribution of peak magnitude for each event. Therefore, according to this definition,it is not deconvolved from the event rate density nor divided by the comoving volume.

\begin{figure*}[htp]
  \includegraphics[width=3.5in]{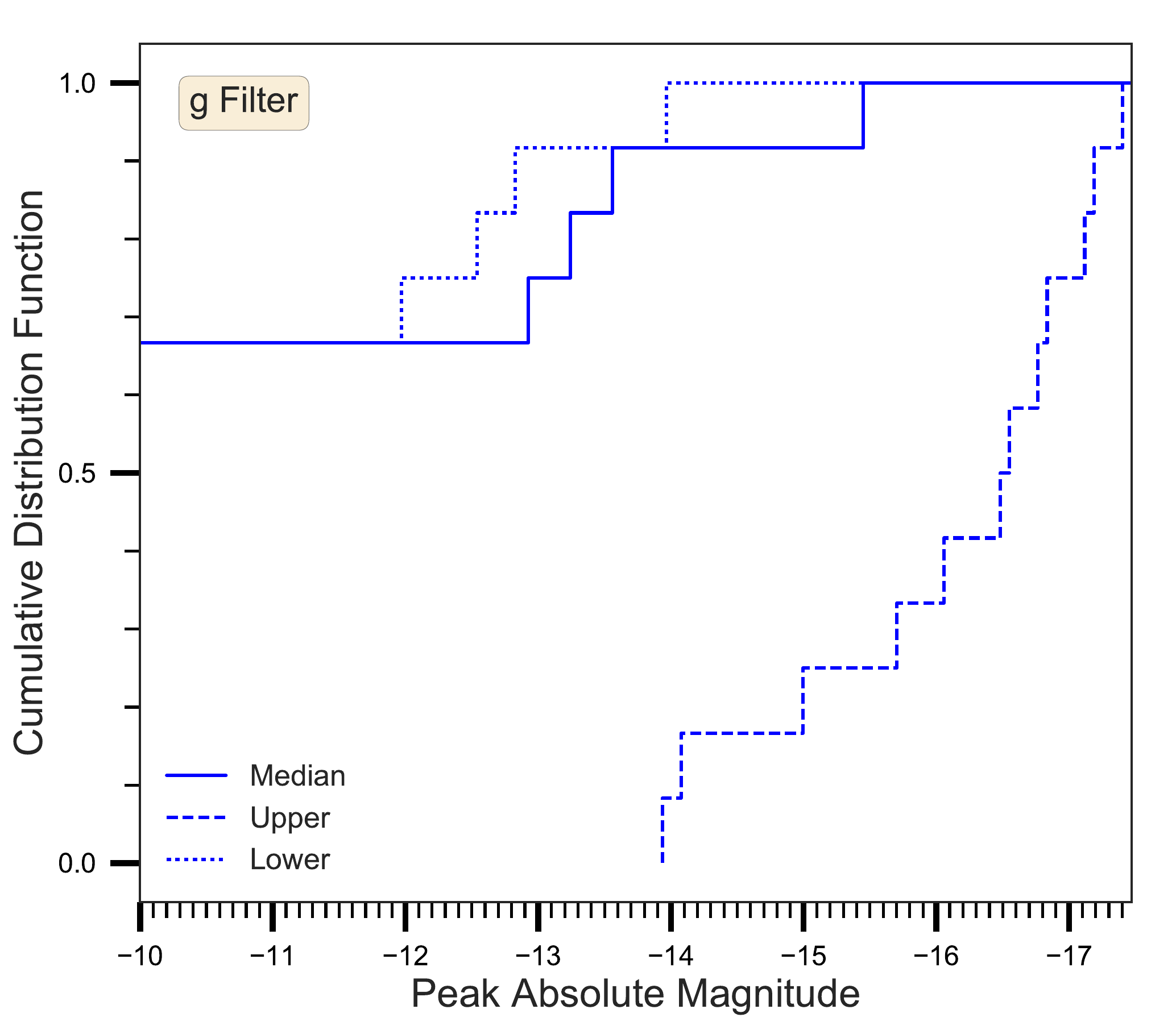}
  \includegraphics[width=3.5in]{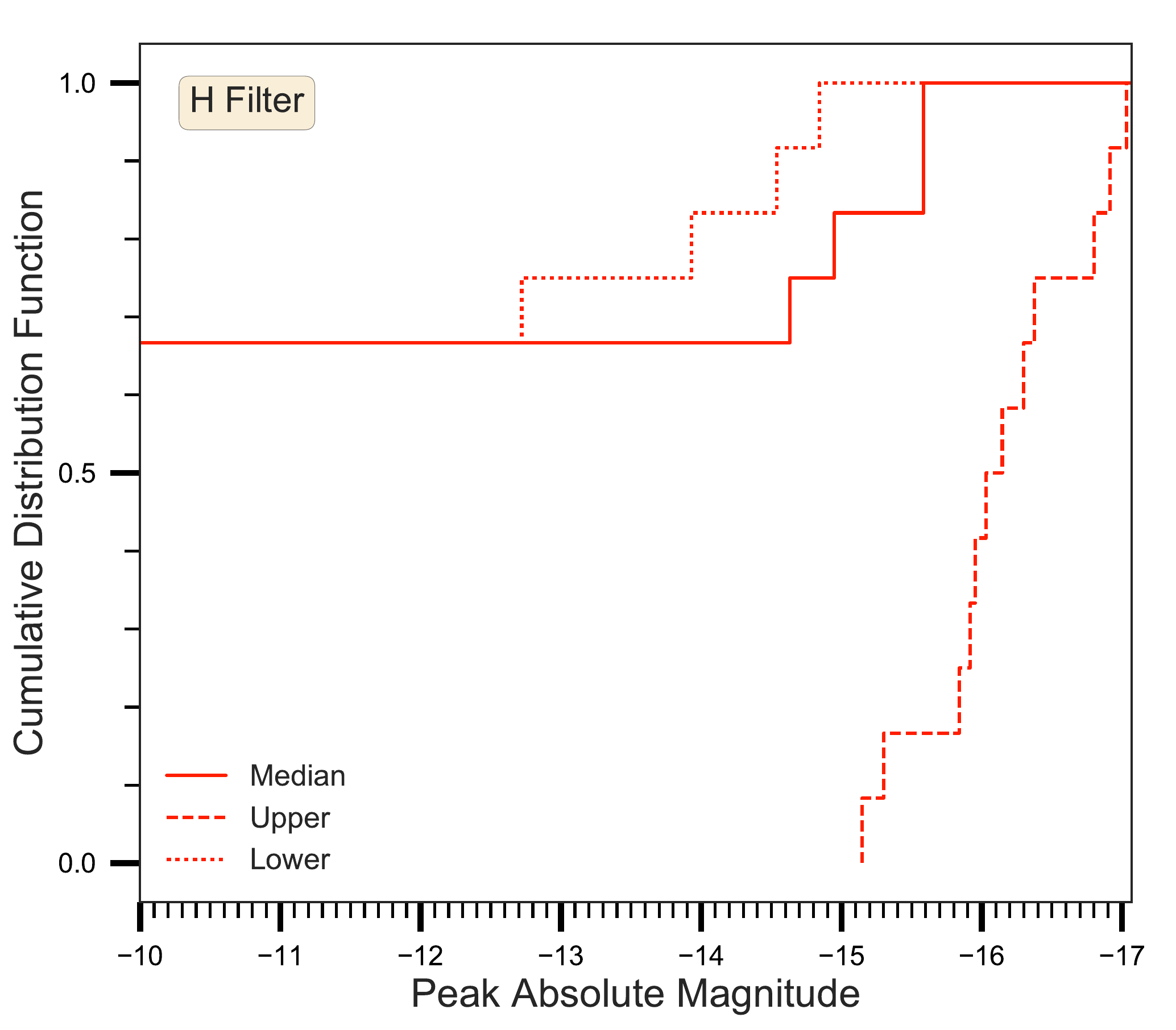}
      \caption{Kilonova luminosity distribution in g (left) and H (right) filters. The solid, dashed and dotted lines are obtained from the median, the upper and the lower limits of the distribution respectively. 
}
 \label{fig:lum_func}
\end{figure*}

\begin{figure*}[htp]
  \includegraphics[width=3.5in]{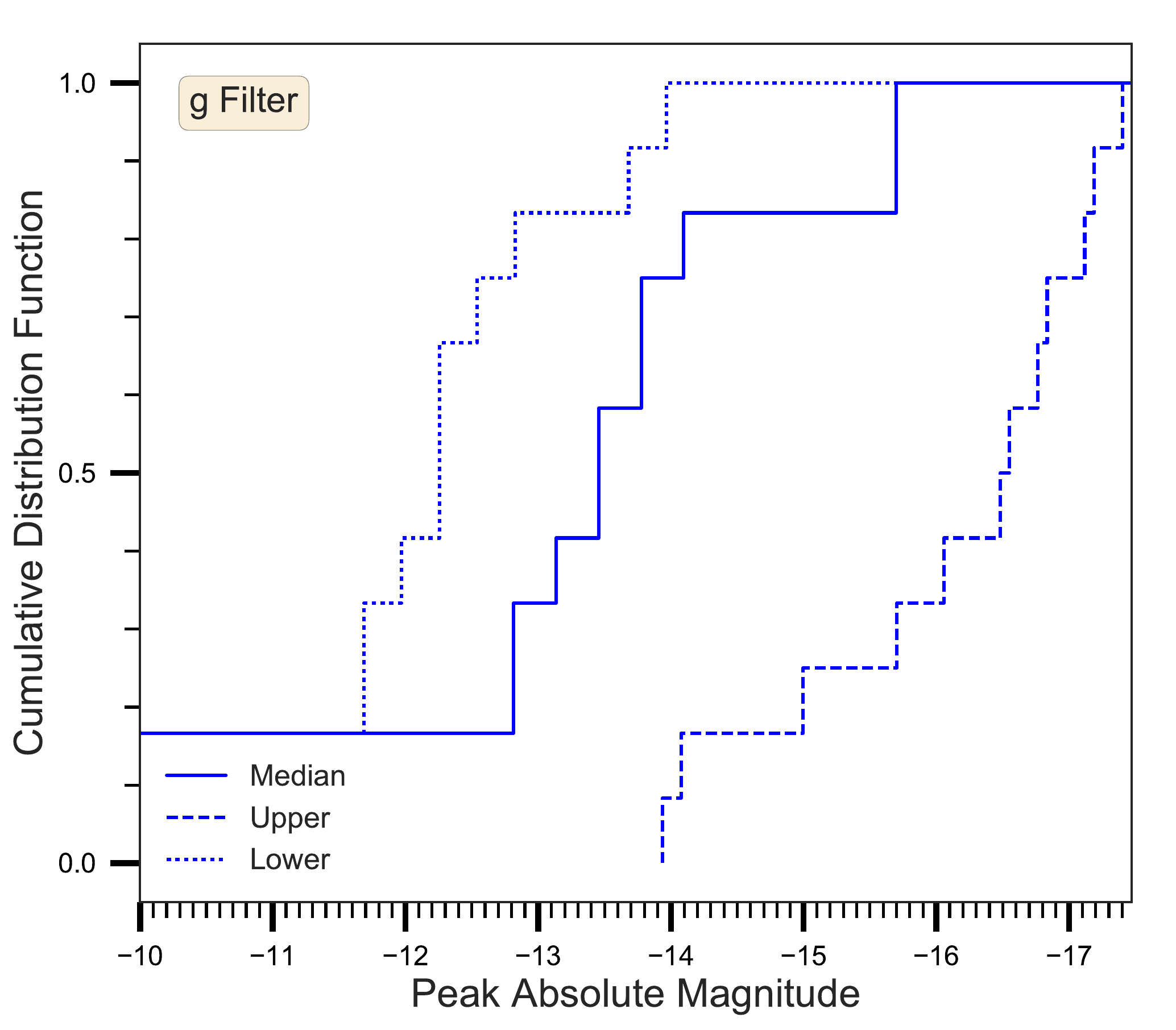}
  \includegraphics[width=3.5in]{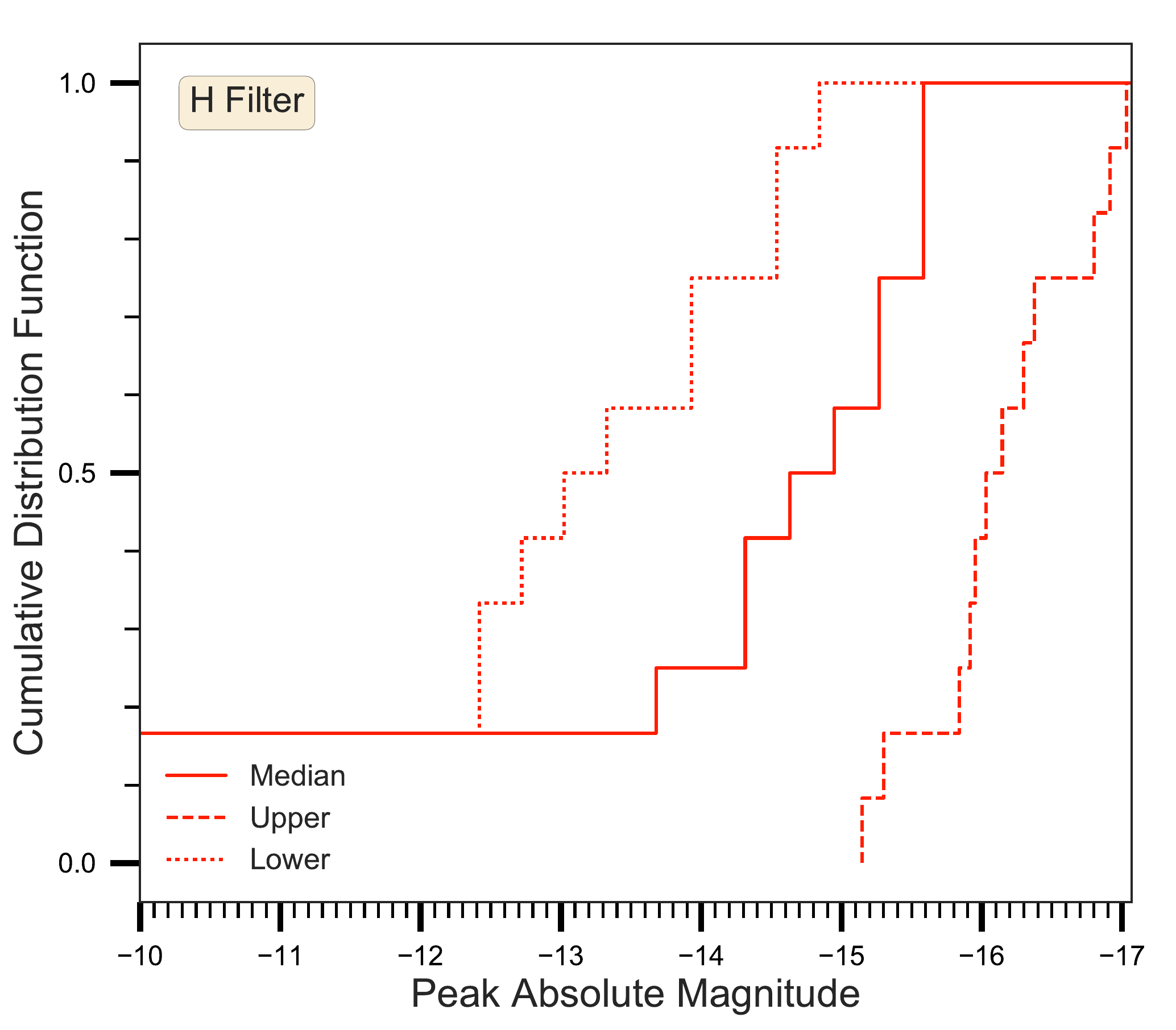}
  \caption{Same as Fig.\ref{fig:lum_func} but here we promoted GRB150101B, GRB050724A, GRB061201, GRB080905A, GRB150424A, GRB160821B to kilonova events.
}
 \label{fig:lum_func_plus}
\end{figure*}

In Fig. \ref{fig:lum_func} we show the luminosity distributions in g and H bands, while the results in all the other bands are reported in Appendix in Fig. \ref{fig:lum_func_allbands}. Due to the recent claim of a kilonova associated to the event GRB150101B and considering also the fact that even in this analysis the lightcurve of this event can be fitted by a kilonova model we decided to repeat the same analysis promoting GRB150101B as a real detection. We promoted also the 5 events found by \citet{Rossi2019}. Since all these claims have been made only recently and the kilonovae differs from the previous kilonovae associated to sGRBs \footnote{GRB150101B is bluer in color, while the claim of a kilonova presence for GRB050724A, GRB061201, GRB080905A, GRB150424A, GRB160821B is due to the observation of an anomalous shallow decay instead of an excess in the afterglow lightcurve.}, we found it useful to include these events separately, in order to allow the comparison of luminosity distributions with and without the inclusion of these events as real kilonova detections. The luminosity distributions obtained with the inclusion of GRB150101B are shown in Fig. \ref{fig:lum_func_plus} and \ref{fig:lum_func_allbands_plus}. 

The luminosity distribution that we obtained allowed us to estimate the exposure time needed for a given telescope facility to detect a kilonova in a given band at a given time. Consider for example the upper limit distribution (dashed line) in Figure \ref{fig:lum_func}. We can see that in g and H bands the $50\%$ of the events are expected to be fainter respectively than $-16.5$ and $-16$ AB absolute magnitude. We can take these values as a benchmark to compute the exposure time needed to detect the 50\% of the events according to optimistic luminosity distribution. If we consider kilonovae at a fiducial distance of $200\,\rm Mpc$ these values translates to an apparent AB magnitude of $20.0$ and $20.5$ respectively. Moreover, in g band the transient is expected to peak within the first $18\,\rm hr$ after the merger/sGRB prompt, while in H band the peak is going to occur between the first and the fourth day. Using the public Exposure Time Calculators (ETC) of GEMINI\footnote{http://www.gemini.edu/sciops/instruments/integration-time-calculators} we estimated an exposure time of $\sim 11.5\,\rm s$ with the instrument GMOS to detect this magnitude in g filter with a signal-to-noise ratio (S/N) $= 20$ within the first day of observation. In the subsequent four days the source could be observed with the instrument NIRI in H filter with 30 exposures each of $\sim8\,\rm s$ (equivalent to a total integration time of $240\,\rm s$) to reach a $\rm S/N = 20\,\rm$.
We repeat the same exercise for VLT\footnote{https://www.eso.org/observing/etc/}. In this case the kilonova can be observed in g band within the first $16\,\rm hr$ using FORS2 imager with a single exposure lasting $\sim 1.2\,\rm s$ (value obtained with input magnitude of 20.0 in B filter as a proxy). In H band instead the transient can be observed with HAWK-I imager with two exposures of $\sim 35\,\rm s$ for a total exposure time of $70\, \rm s$. For both GEMINI and VLT a typical airmass of $1.5$ has been considered in this calculation.

\subsection{Local density of r-process elements} 

 \begin{figure*}[t]
 \centering
 \includegraphics[width=6.0in]{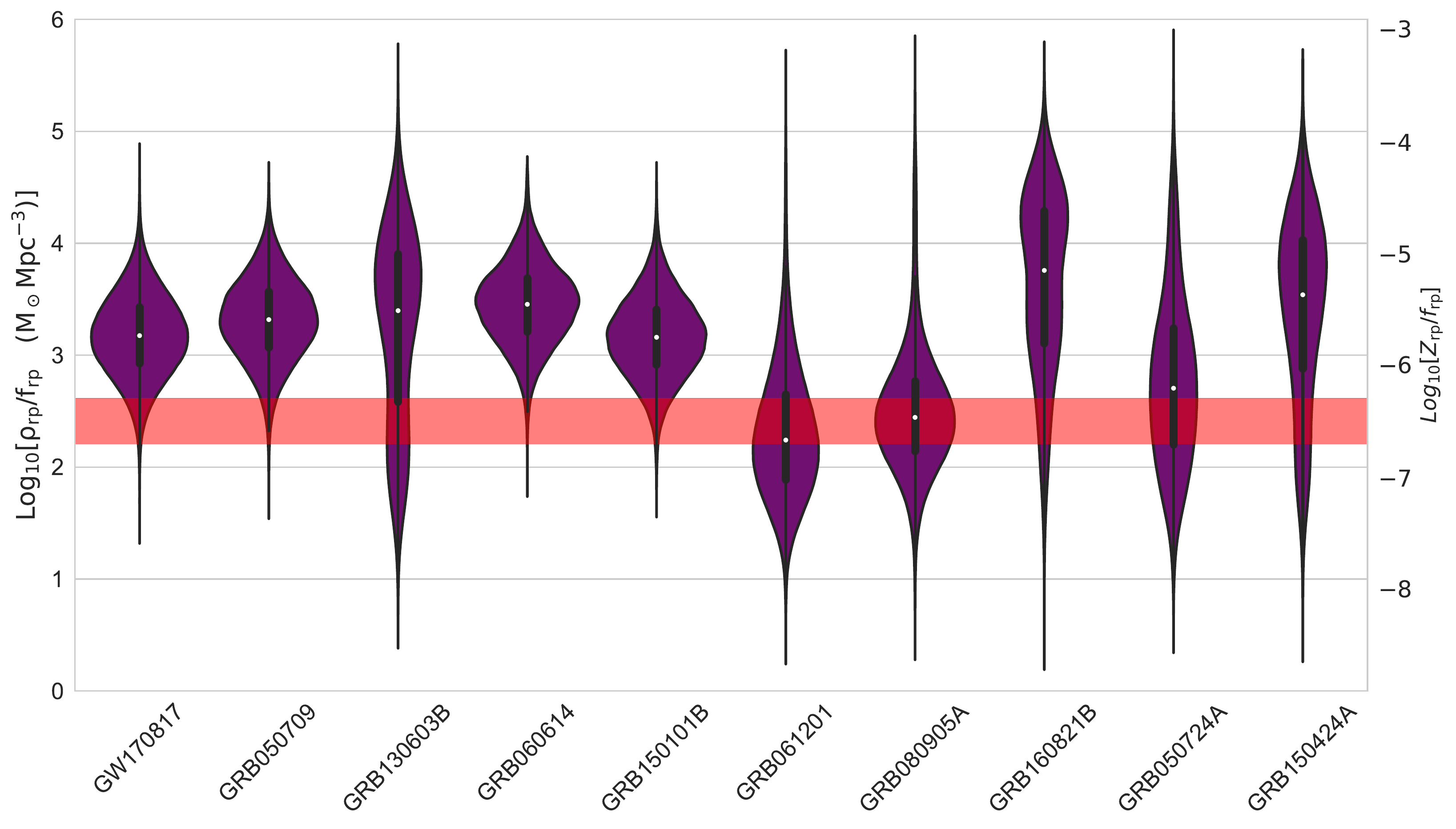}
 \caption{Distributions of average local r-process elements density and fraction estimated for the events  GW170817, GRB050709, GRB151010B, GRB130603B, GRB060614, GRB061201, GRB080905A, 160821B, GRB050724A and GRB150424A. The red band represents r-process mass fractions obtained from Solar system measurements \citep{Arnould2007}.}
 \label{fig:local_density}
\end{figure*}

Finally we repeat the analysis of \citet{AbEA2017f} to estimate the average dynamically ejected local r-process material density $\rho_{\rm rp}$ for the events GW170817, GRB050709, GRB151010B, GRB130603B, GRB060614 and GRB050724A,  GRB061201,  GRB080905A, GRB150424A,  GRB160821B. The average local density is calculated according to the formula:

\begin{equation}
	\rho_{\rm rp}/f_{\rm rp} = M_{\rm ej}\mathcal{R}\frac{\int^{t_H}_0 \int^t_0\dot{\rho}_*(\tau)p_{\rm delay}(t-\tau)d\tau dt}{\int^{t_H}_0 \dot{\rho}_*(\tau)p_{\rm delay}(t_H-\tau)d\tau}
\end{equation}
 where $f_{\rm rp}$ is the fraction of dynamical ejecta matter converted in r-process elements, $\mathcal{R}$ is the present day merger rate, $t_H$ is the Hubble time, $\dot{\rho}_*(t)$ is the star formation rate of \citet{MaDi2014}, $p_{\rm delay}\propto 1/t$ is the distribution of delay time between the BNS formation and its merger \citep{OsBel2008,Dominik2012}. As in \citet{AbEA2017f} integrating over the cosmic history a $\rm \Lambda CDM$ cosmology with parameters in \citet{Planck2016} have been assumed. 
Here we sample from $M_{\rm ej}$ distributions for the single event, while for the present day rate we sample over a log-normal distribution with a $90\%$ confidence in the range $[360, 4730]\,\rm Gpc^{-3}\rm yr^{-1}$ (consistent with the rates inferred from GW170817 \citep{AbEA2017b}. 
Our results are reported in Figure \ref{fig:local_density}, where we report on the left side also the mass fraction of r-process elements calculated as $Z_{\rm rp}/f_{\rm rp} \equiv (\rho_{\rm rp}/f_{\rm rp})/\rho_*$, with $\rho_* = \int^{t_H}_0\dot{\rho}_*dt$. 
It is worth noting that the fractions $X_{\rm lan}$ and $Z_{\rm rp}$ denote two different quantities: $X_{\rm lan}$ is the lanthanide fraction in the merger ejecta, while $Z_{\rm rp}$ is the average mass fraction of r-processes all elements (lanthanides included) in the present day universe, calculated assuming that all the BNS mergers contribute to the enrichment with the $M_{\rm ej}$ of the given event. 

The red band here denotes the r-process elements mass fraction from Solar system observations \citep{Arnould2007}.

It is worth noticing that the average local density obtained from GW170817 is about an order of magnitude higher than that obtained by \citet{AbEA2017f}. This discrepancy is due to the different ejecta mass distribution employed in this work (obtained from lightcurve fitting), that is about one order of  magnitude higher than that used in \citep{AbEA2017f} (obtained from the BNS masses distributions plus \citep{DiUj2017} fitting formula). Nevertheless our results are still (marginally) consistent with the Solar system measured mass-fractions and illustrate the unceratinities associated with deriving accurate ejecta masses given primarily our lack of understanding of r-process  opacities.  In all cases, our masses are above the stringent minimum mass requirements derived from low metallicity stars in the Universe \citep{ShCoRR2015, MaRR2018}.

\section{Summary}
\label{sec:summary}

Our analysis is the first study using the latest models of AT2017gfo and presents posteriors of $M_{\rm ej}$, $v_{\rm ej}$ and  $X_{\rm ej}$ for the ''historical'' sGRBs. 
In general, both the absolute magnitude predictions and the color evolution of the kilonovae allow for the differentiation of their contribution from the afterglow.

While GRB130603B was the first short GRB with evidence for a kilonovae followed by GRB050709 and GRB060614, other short GRBs (GRB061201, GRB080905A and marginally GRB160821B) provide constraints on r-process rich ejecta contributions to those lightcurves. 
GW170817, the first joint GW-EM detection, provides tighter constraints than GRB130603B both in peak time and luminosity and we expect to observe more events like this in the near future. 
GRB130603B and GW170817 are in our sample the real kilonova detections, for which our analysis predicts a dominant H over g filter luminosity. In the other events that provide upper limits the H luminosity is dominant as well, with the only exception of GRB150101B. 

Considering both real detections and upper limits our analysis identify so far an H filter peak magnitude in the range of $[-16.2,-13.1]$ (along with a H band peak time in range $[0.8,3.6]$ days). 
We use our sample of nearby ($z < 0.5$) sGRBs which comprise both events with and without a kilonova candidate to draw the first kilonova luminosity distribution in literature in different frequency bands. We build three different limiting luminosity distributions corresponding to the median, lower and upper limiting values of the peak luminosity distributions. Our results obtained considering the kilonova candidates GRB130603B, GRB050709 and GRB060614 as real kilonovae (Fig. \ref{fig:lum_func}) show that in the H filter half of the events are below the -16th mag in the optimistic case (dashed line) while in the median (solid line) and pessimistic case (dotted line) about the 64\% of the events are below the -14.6 mag and -12.7 mag respectively. Including GRB150101B, GRB050724A,  GRB061201,  GRB080905A, GRB150424A,  GRB160821B  as real kilonovae reduces the difference between the lower and median distribution with the upper distribution, which is in fact a consequence of the fact that in our sample we have more upper limits than real kilonovae events. In this case the median and the lower distributions result in about 17\% of the events below -13.8 mag and -12.4 mag respectively, while the upper limit is unchanged by construction. 
For the luminosity distribution in g filter we observe that the upper case predict half of the events fainter than -16.3 mag, while the lower and the median case predict the 67\% of events below the -12.9 mag and -11.7 mag respectively. Including GRB150101B, GRB050724A,  GRB061201,  GRB080905A, GRB150424A,  GRB160821B we obtain that 17\% are fainter than -12.8 mag and -11.7 mag in the median and lower distributions respectively. We expect that future observations of kilonovae will help to reduce the uncertainties between the three limiting distributions considered here.

The results obtained in the present work can be used to predict the absolute magnitude and color of future events, and inform the search strategies that will be used to detect them. 
This could include using the predictions and the three-dimensional skymaps to allocate exposure times sufficient to make detections (see for example \citet{SalafiaColpi2017} and \citet{CoughlinAhumada2019}).
In this way, the kilonova detections can be used as benchmarks for future searches.
Further statistical samples will enable making constraints on the progenitor system properties, including the mass ratio and equation of state, based on the lightcurves alone \citep{CoDi2018b}.
Moreover, under the assumption that all BNS are the progenitors of sGRBs, the mergers can be used to constrain their overall contribution to the r-process in the universe \citep{AbEA2017f}.

Future observations, coupled with more detailed theory models, will allow us to place more stringent constraints on the kilonova peak luminosity distribution. Moreover they will allow us to answer the following questions: 
\begin{itemize}
\item Are the kilonovae produced by NS-NS mergers different from those produced by NS-BH mergers (if any)?
\item Are the kilonovae produced in NS-NS merger events with BH remnant different from those produced in merger events with a NS remnant ? 
\item How does the binary system inclination angle influence the kilonova characteristics (color, peak luminosity) ? 
\end{itemize}

Theory predicts that the nature of the progenitor and the merger remnant along with inclination angle of the binary could have an impact on the observable feature of the transient \citep{RoKa2011}.
Hydrodynamical simulations show for example that a NS-BH merger is expected to dynamically eject more mass than a NS-NS coalescence \citep{Rosswog2015}, thus generating a more luminous transient. On the other hand, during a BNS merger a part of the ejecta is the result of shocks that emerge from the contact interfaces between the stars. This matter component reaches large enough temperatures ($\sim \rm MeV$) to undergo fast positron captures and can thus reach electron fractions that substantially differ from the original, very low beta-equilibrium values. The same is true
   if a massive NS survives at least temporarily the merger event. In this
   case, strong neutrino-driven winds emerge with a range of electron fractions
   from $Y_e \sim 0.2 - 0.4$ \citep{PeRo2014}. For both types of ejecta --shock- or neutrino-driven-- a substantial mass
   fraction is above the critical value $Y_e^{\rm crit}= 0.25$ above which no
   more lanthanides are produced \citep{KoRo2012}. Therefore the resulting
   transients are blue. These components, if present, are ejected
mainly perpendicularly to the orbital plane inducing in this way a viewing angle dependence \citep{WaSe2014,WoKo2017}.
The angular dependence may reflect in a shift of the kilonova luminosity distribution towards higher magnitudes in the optical bands. This could be verified in the near future, when kilonovae at larger viewing angle will be likely observed in association with GW events.
Furthermore a BNS coalescence could result in the formation of a highly magnetized fast spinning NS, which can be either stable or centrifugally supported by rigid or differential rotation and undergo a delayed collapse into a BH \citep{GP13,GiZr2015,FrBel2015, CiKa2017,PiGi2017,RaPe2018}. In this scenario the NS dipole spindown emission would constitute an additional source of energy that would heat the ejecta and boost its expansion resulting, once again in different observational features (which would depend also on the remnant NS parameters) \citep{YZG13,GD13, MP14,SC16a,SC16b}. On the other hand, the presence of the accompaying neutrino driven wind might prevent the emergence of a sGRB \citep{MuMo2014, MuRR2017a}. In this scenario we could expect the presence of transients with higher luminosities and a spectrum peaked at higher energies not associated with sGRBs, either due to the orientation of the observer or the hampering of the jet. In the case this scenario occurs in a substantial fraction of BNS merger -and considering these sources as proper kilonovae- we could expect a shift of the luminosity distribution towards lower magnitudes in (at least) the optical filters.

We are now seeing a renaissance in both the ejecta and light-curve models from kilonova.  Improvements in theory are eliminating or placing constraints on uncertainties in the nuclear heating, atomic opacities and transport methods.  In addition, a better understanding of the ejection properties are producing more physical ejecta profiles that will lead to more accurate ties between emission and ejecta masses.  With these models, the electromagnetic detections will provide a tight connection to the properties of the mergers.  Combined with GW detections, these observations will be able to assess the nature of the progenitor and the merger remnant and measure the viewing angle.  These join GW-electromagnetic observations will also place constraints on the role of shocks in the afterglow emission. Assessing whether different remnants lead to different kilonova events is important because, if true, would allow to identify the nature of the remnant from the only electromagnetic emission, therefore even in case of poor signal to noise ratio of GW post-merger signal.  

\acknowledgments

This work was initiated
and supported by the 2017 Kavli Summer Program
in Astrophysics at the Niels Bohr Institute in Copenhagen,
and the authors would like to thank DARK at the University
of Copenhagen for incredible hospitality. The 2017
Kavli Summer Program program was supported by the the
Kavli Foundation, Danish National Research Foundation
(DNRF), the Niels Bohr International Academy and DARK.

A.M.B.\ acknowledges support from a UCMEXUS-CONACYT Doctoral Fellowship.

M.C.\ is supported by the David and Ellen Lee Postdoctoral Fellowship at the California Institute of Technology.

T.D.\ acknowledges support by the European Union’s Horizon  2020 research and innovation program under grant agreement No 749145, BNSmergers.

R.J.F.\ is supported in part by NASA grant NNG17PX03C, NSF grant AST-1518052, the Gordon \& Betty Moore Foundation, the Heising-Simons Foundation, and by fellowships from the Alfred P.\ Sloan Foundation and the David and Lucile Packard Foundation.

E.R.-R.\ is supported in part by David and Lucile Packard Foundation and the Niels Bohr Professorship from the DNRF.  

S.A. acknowledges Giulia Stratta, Daniele Malesani, Andrea Melandri and Enzo Brocato for useful discussions.

The authors acknowledges the anonymous referee for his/her suggestions that help us to improve the presentation of our results.

\bibliographystyle{aasjournal}
\bibliography{references}

\newpage 

\appendix
\renewcommand\thefigure{\thesection A.\arabic{figure}} 
\renewcommand*\thetable{\Alph{section}A.\arabic{table}}

\setcounter{figure}{0}  

\begin{table}[t]
\centering
\setlength{\tabcolsep}{10pt}
	\begin{tabular}{cccccc}
		\toprule
        GRB & Ref.  & Kilonova & $M_{\rm ej}\quad(\rm M_\odot)$  &$X_{\rm lan}$ & Redshift \\
        \midrule
        
        170817A (GW170817) & 1 & yes & $3.87^{+3.39}_{-1.44}\times 10^{-2}$ & $2.71^{+8.60}_{-2.03}\times 10^{-4}$ & 0.0099\\[4ex]
        130603B & 2, 3  & yes & $7.46^{+43.97}_{-7.29} \times 10^{-2}$ & $5.36^{+64.63}_{-5.36} \times 10^{-3}$ & 0.356\\[4ex]
        050709 & 4, 5, 6, 7 & yes & $5.11^{+2.98}_{-2.13} \times 10^{-2}$ & $4.49^{+49.60}_{-4.45}\times 10^{-5}$ & 0.161\\[4ex]
060614 & 8, 9, 10 & yes & $7.73^{+1.90}_{-2.85} \times 10^{-2}$ & $2.24^{+36.73}_{-2.23}\times 10^{-6}$ & 0.125\\[4ex]
        150101B & 11, 12  & recently claimed by 12 & $3.71^{+3.12}_{-1.56}\times 10^ {-2}$ & $4.19^{+889.60}_{-4.16} \times 10^{-7}$ & 0.134\\[4ex]

        140903A & 13 & no & - & - & 0.351\\[4ex]
        050724A & 14, 15 & recently claimed by 24 & $1.24^{+39.99}_{-1.09}\times10^{-2}$ & $0.09^{+227.56}_{-0.09}\times 10^{-4}$ & 0.257\\[4ex]
        061201 & 16 & recently claimed by 24 & $4.20^{+38.34}_{-2.91}\times 10^{-3}$& $0.07^{+361.50}_{-0.07}\times 10^{-4}$ & 0.111\footnote{This event have been associated to a galaxy at the redshift reported here or to the cluster Abell 995 at $z=0.084$. \citet{GoLe2018} employed for this event the latter value, while we choose the former in order to be more conservative}\\[4ex]
        080905A & 17, 18 & recently claimed by 24 & $6.98^{+44.01}_{-4.58}\times10^{-3}$& $1.41^{+200.38}_{-1.41}\times 10^{-4}$ & 0.1218\\[4ex]
        070724A & 19, 20 & no & -& - & 0.457\\[4ex]
        160821B & 21, 22 & recently claimed by 24 & $1.74^{+6.97}_{-1.69}\times 10^{-1}$& $1.87^{+175.29}_{-1.87}\times 10^{-4}$ & 0.16\\[4ex]
        150424A & 22, 23 & recently claimed by 24 & $9.66^{+56.04}_{-9.45}\times 10^{-2}$ & $0.15^{+188.15}_{-0.15}\times 10^{-4}$ & 0.30\\[4ex]
        \bottomrule
	\end{tabular}
    \caption{Summary of all GRBs in our samples. Ejecta mass and lanthanide fraction are given for the events with a confirmed kilonova detection and for those with informative upper limits are provided. The reported uncertainties correspond to a 90\% of confidence interval. References: 
(1) \citet{LVC2017}, (2) \citet{TaLe2013}, (3) \citet{BeWo2013}, (4) \citet{FoFr2005}, (5) \citet{Hjorth2005}, (6) \citet{Covino2006}, 
(7) \citet{Jin2016}, (8) \citet{ZhZh2007}, (9) \citet{JinLi2015},
(10) \citet{YaJi2015}, (11) \citet{FoMa2016}, (12) \citet{TrRy2018},
(13) \citet{TrSa2016}, (14) \citet{BePr2005}, (15) \citet{MalesaniCovino2007}
(16) \citet{Stratta2007}, (17) \citet{NicuesaGuelbenzu2012}, 
(18) \citet{RoWi2010}, (19) \citet{BeCeFo2009}, (20) \citet{Kocevski2010}, (21) \citet{KaKoLau2017}, (22) \citet{JinWang2018},
     (23) \citet{TanLevFru2015}, (24) \citet{Rossi2019}}
    \label{tab:all_grbs}
\end{table}

\begin{figure*}[t]
     \centering
    \includegraphics[width = 0.9\columnwidth]{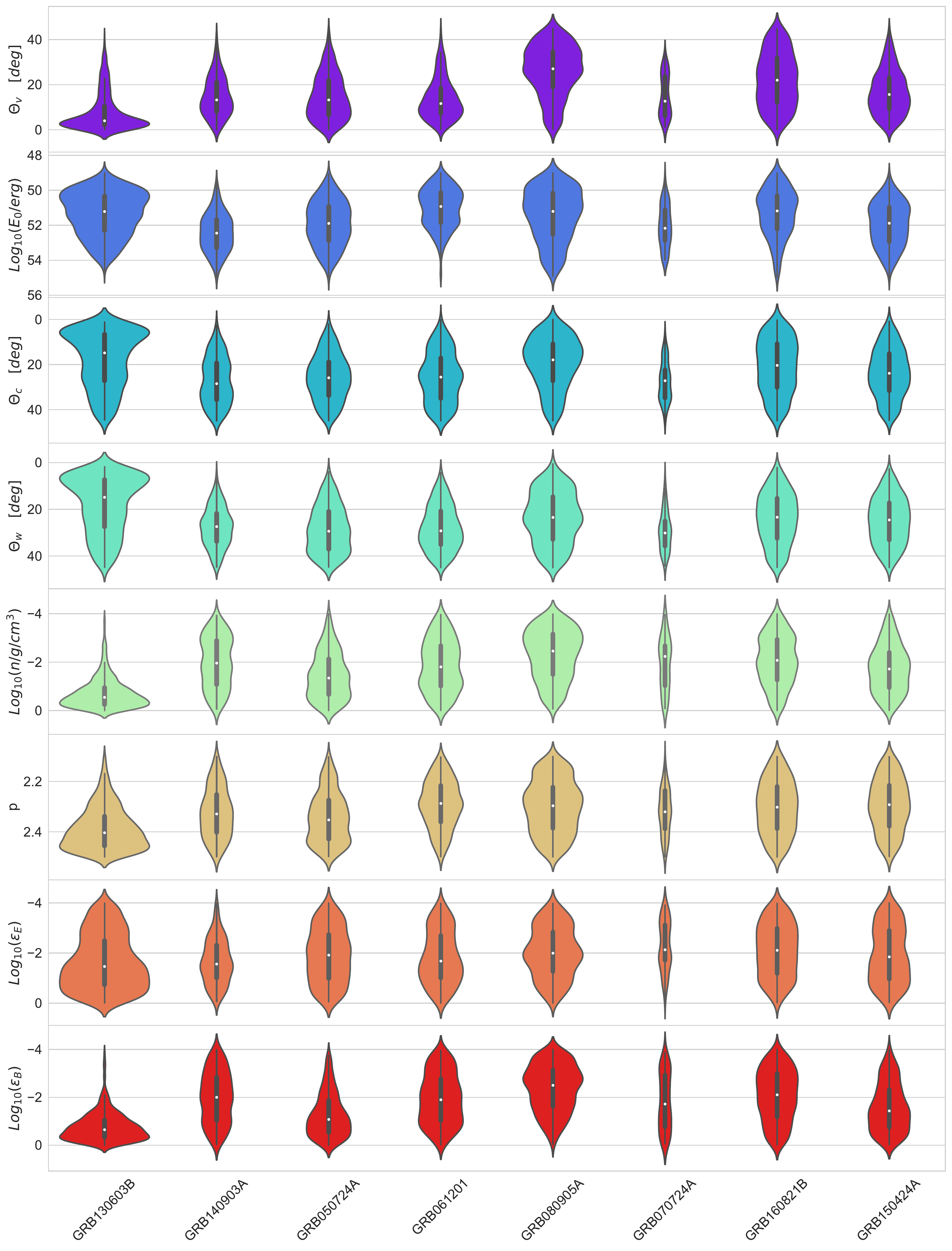}
    \caption{Distributions of the afterglow parameters obtained for the events fitted by the afterglow+kilonova models.}
    \label{fig:afterglow_params}
\end{figure*}

\begin{figure*}[t]
\includegraphics[width=3.5in]{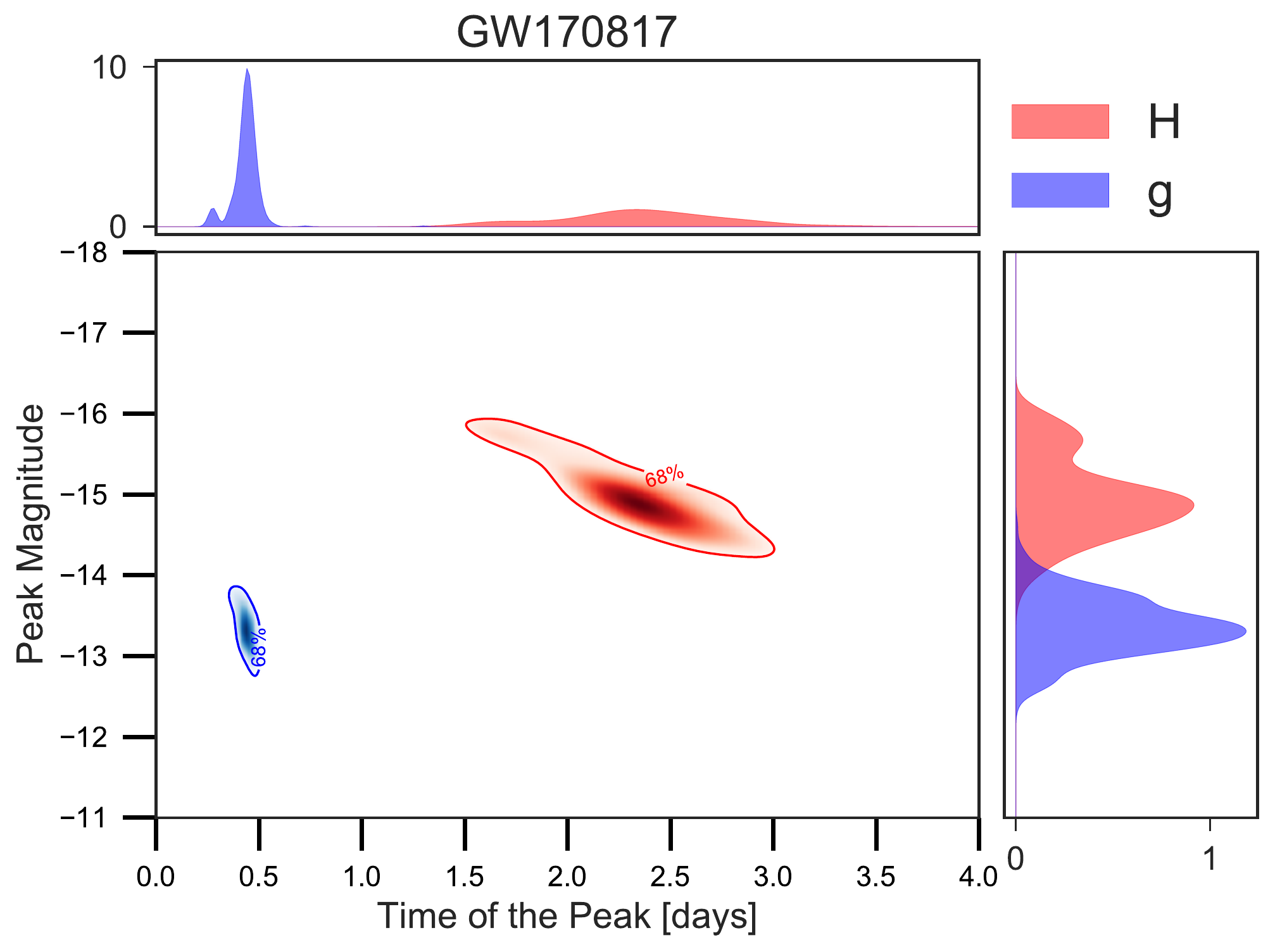}
\includegraphics[width=3.5in]{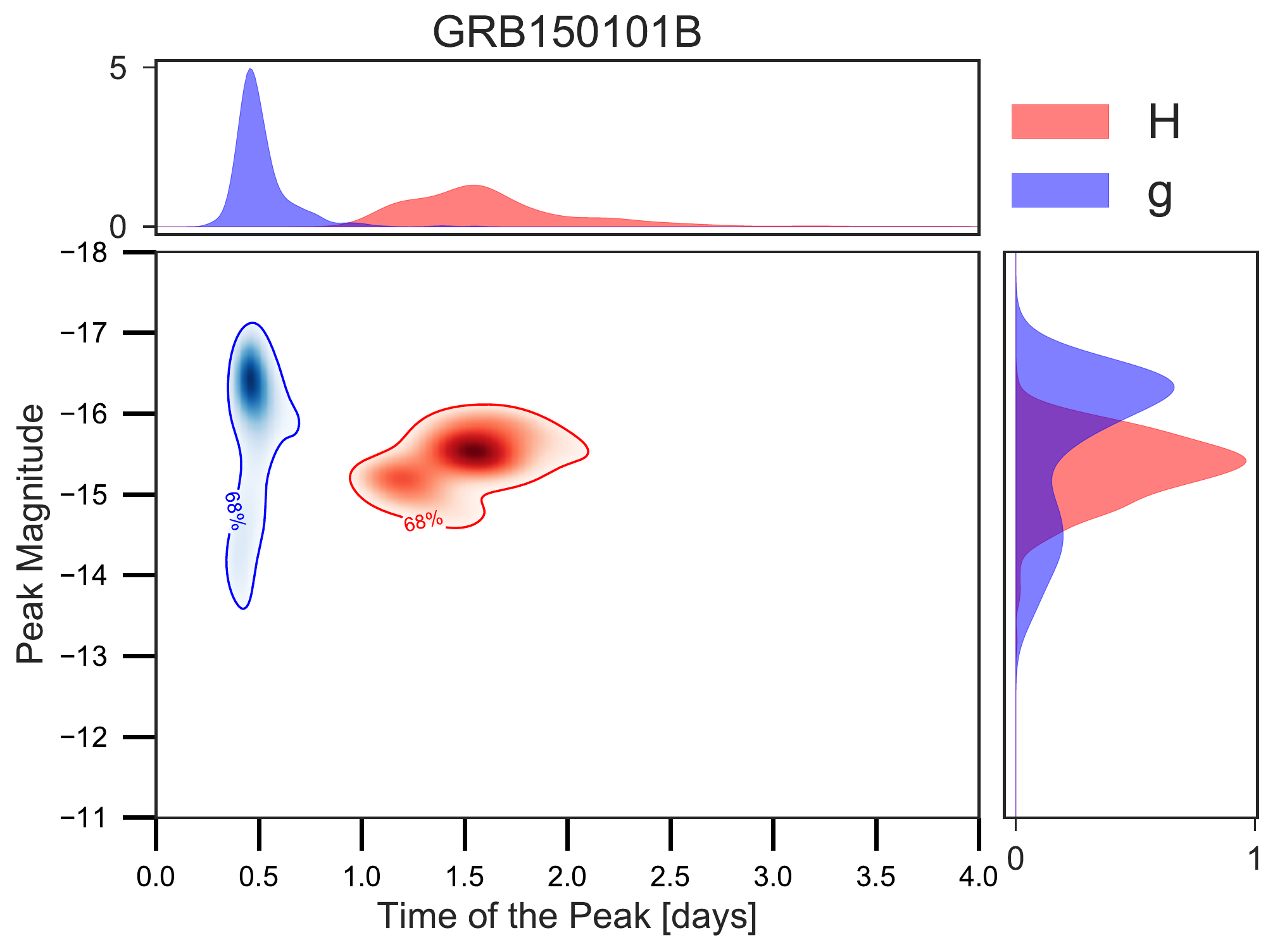}
\includegraphics[width=3.5in]{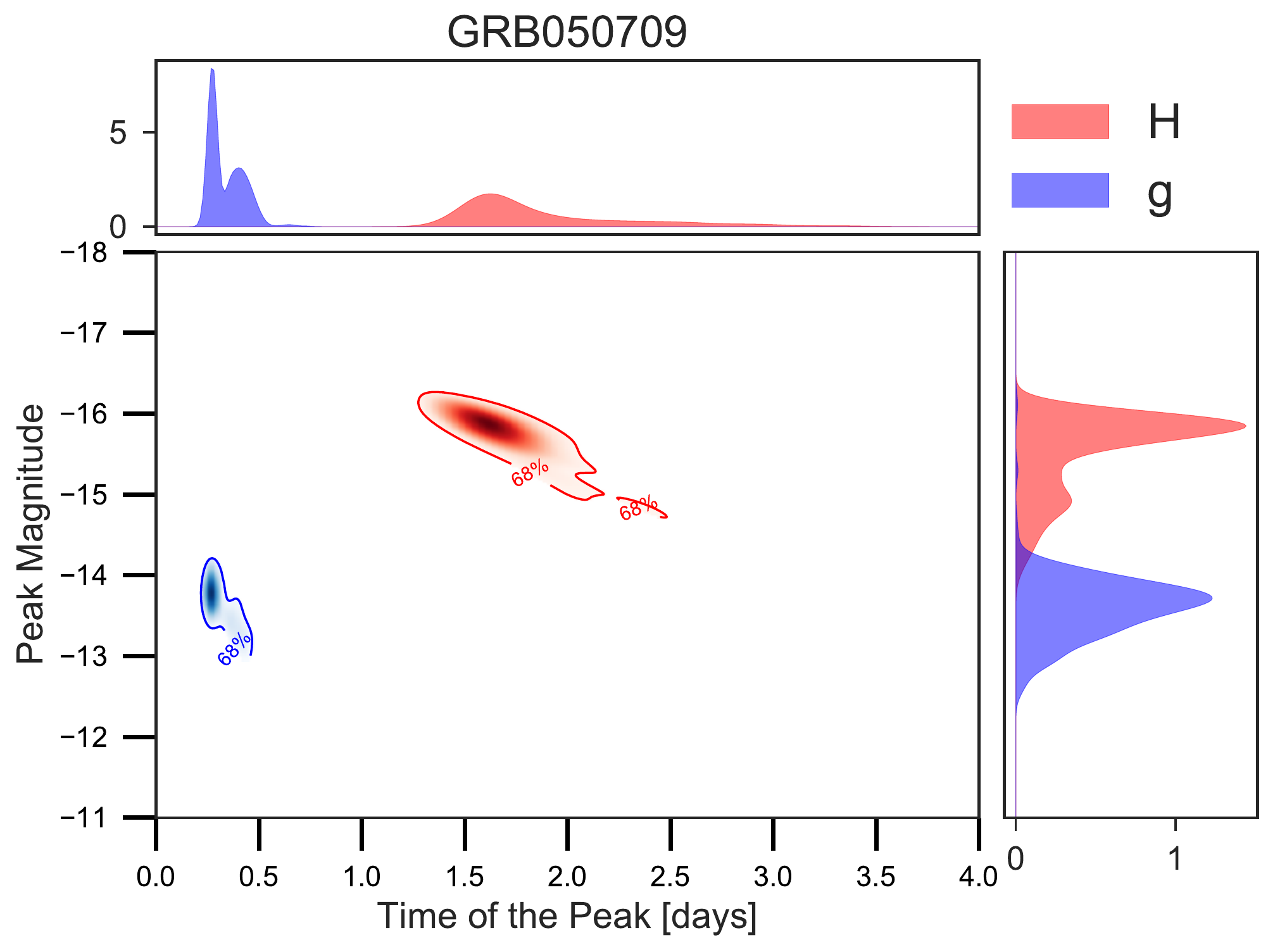}
\includegraphics[width=3.5in]{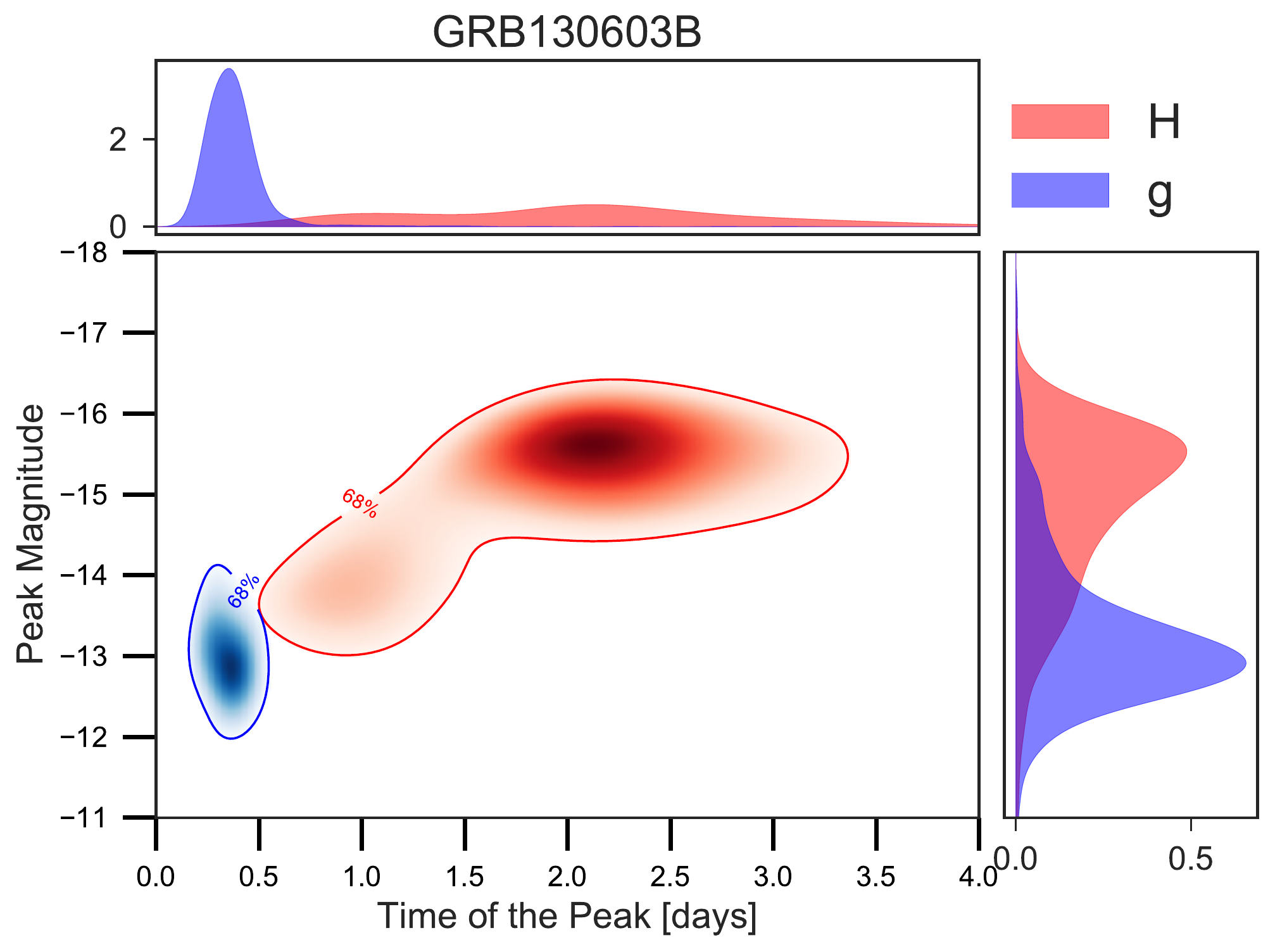}
\centering
\includegraphics[width=3.5in]{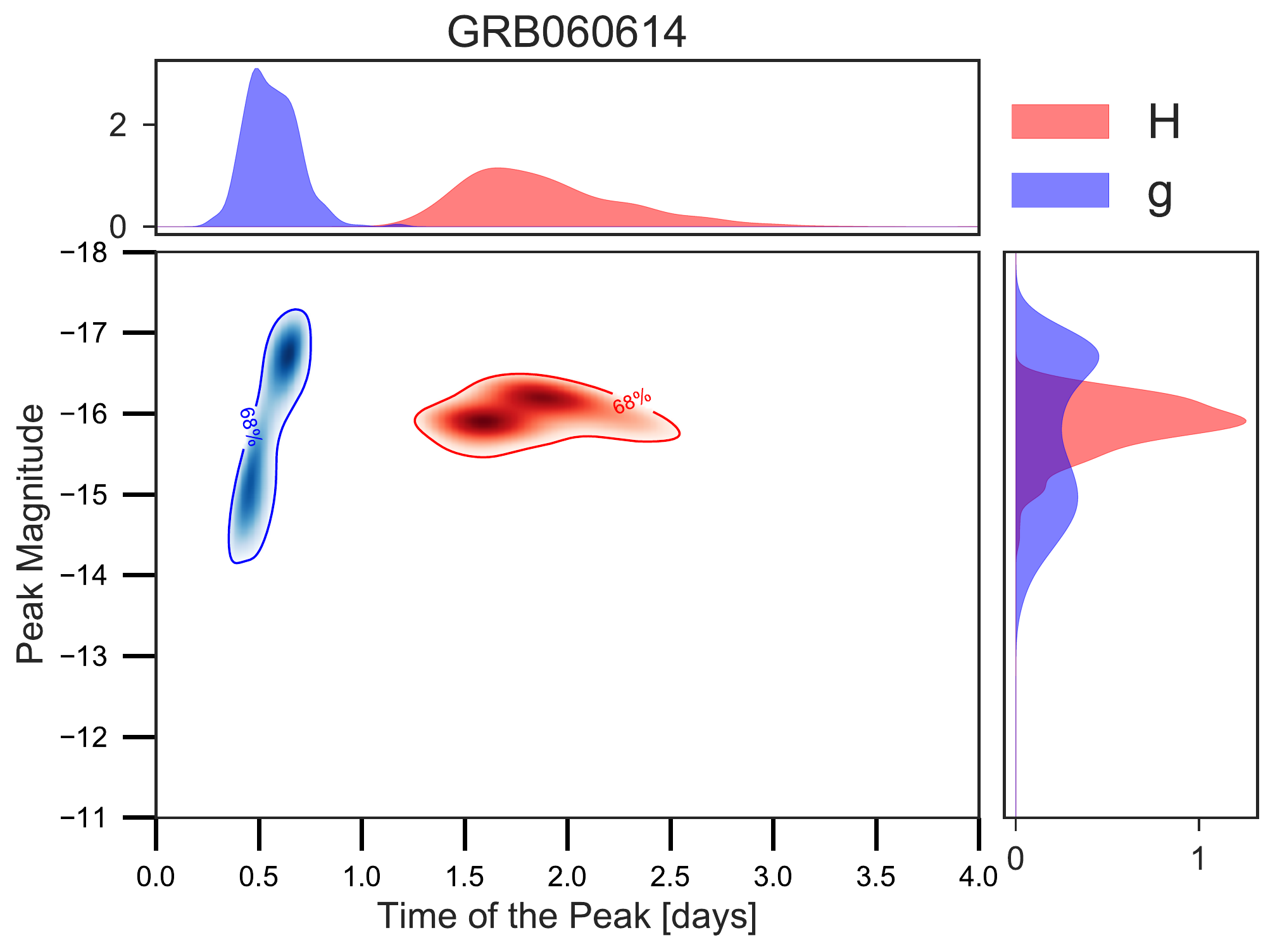}
\caption{Probability density distribution (within 68\% of confidence) in a peak time- peak Magnitude plane in g (blue) and H (red) filters for the events GW170817, GRB150101B, GRB050709, GRB130603B and GRB060614. Top and right panels show the marginalized distributions in peak magnitude and peak time respectively.} 
\label{fig:PeakTimeMag}
\end{figure*}

\begin{figure*}[t]
	\includegraphics[width=3.3in]{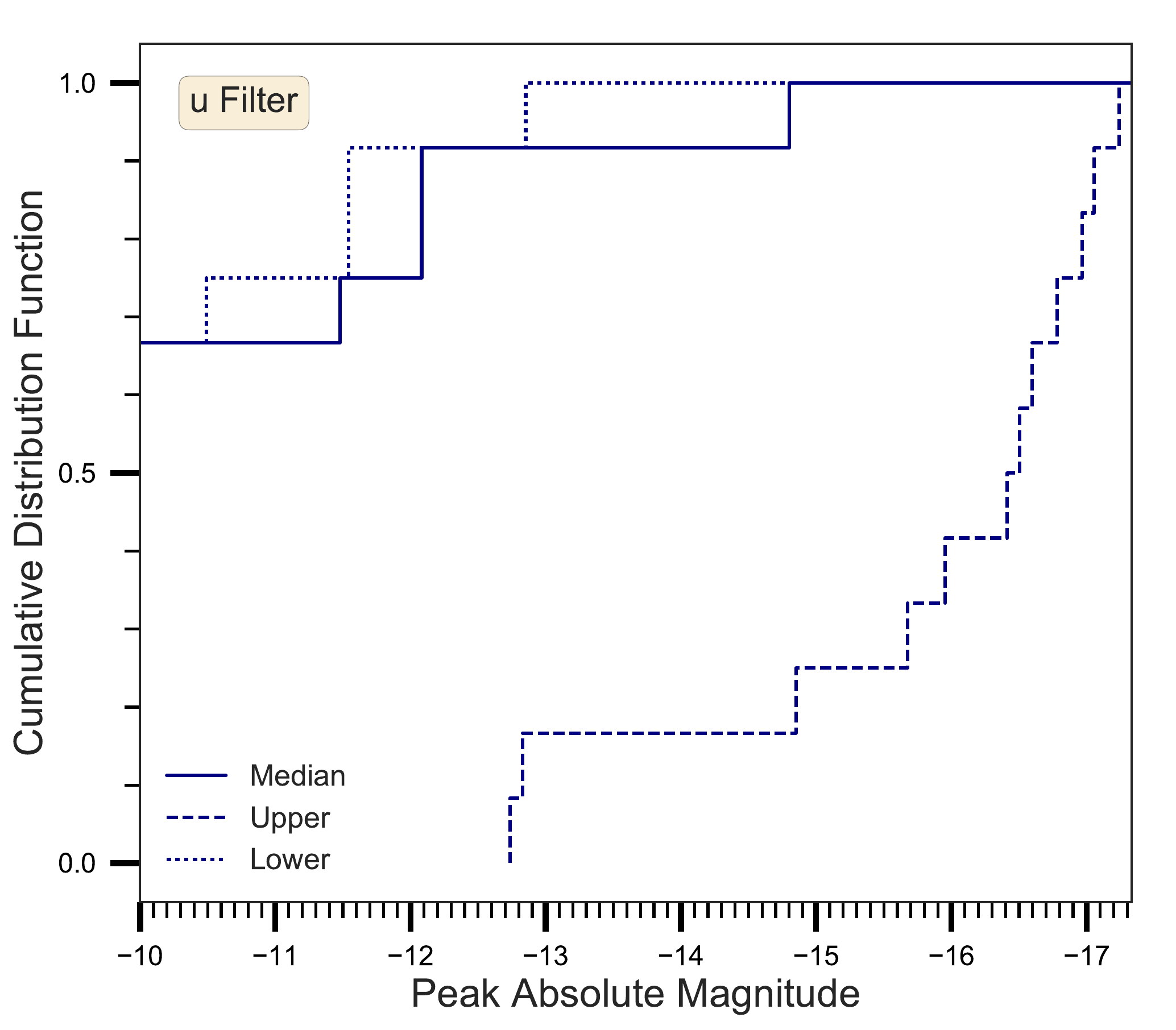}
    \includegraphics[width=3.3in]{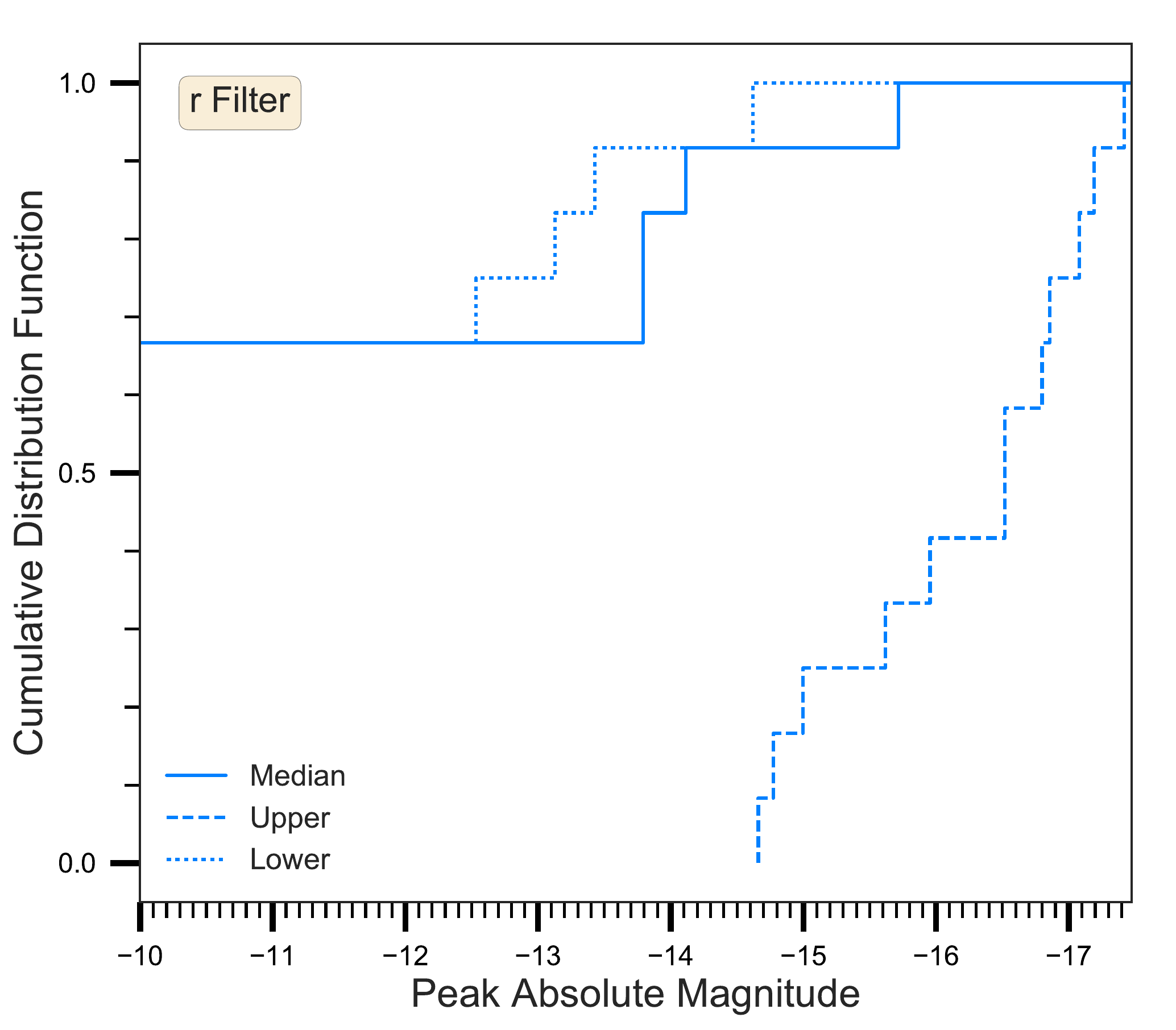}
    \includegraphics[width=3.3in]{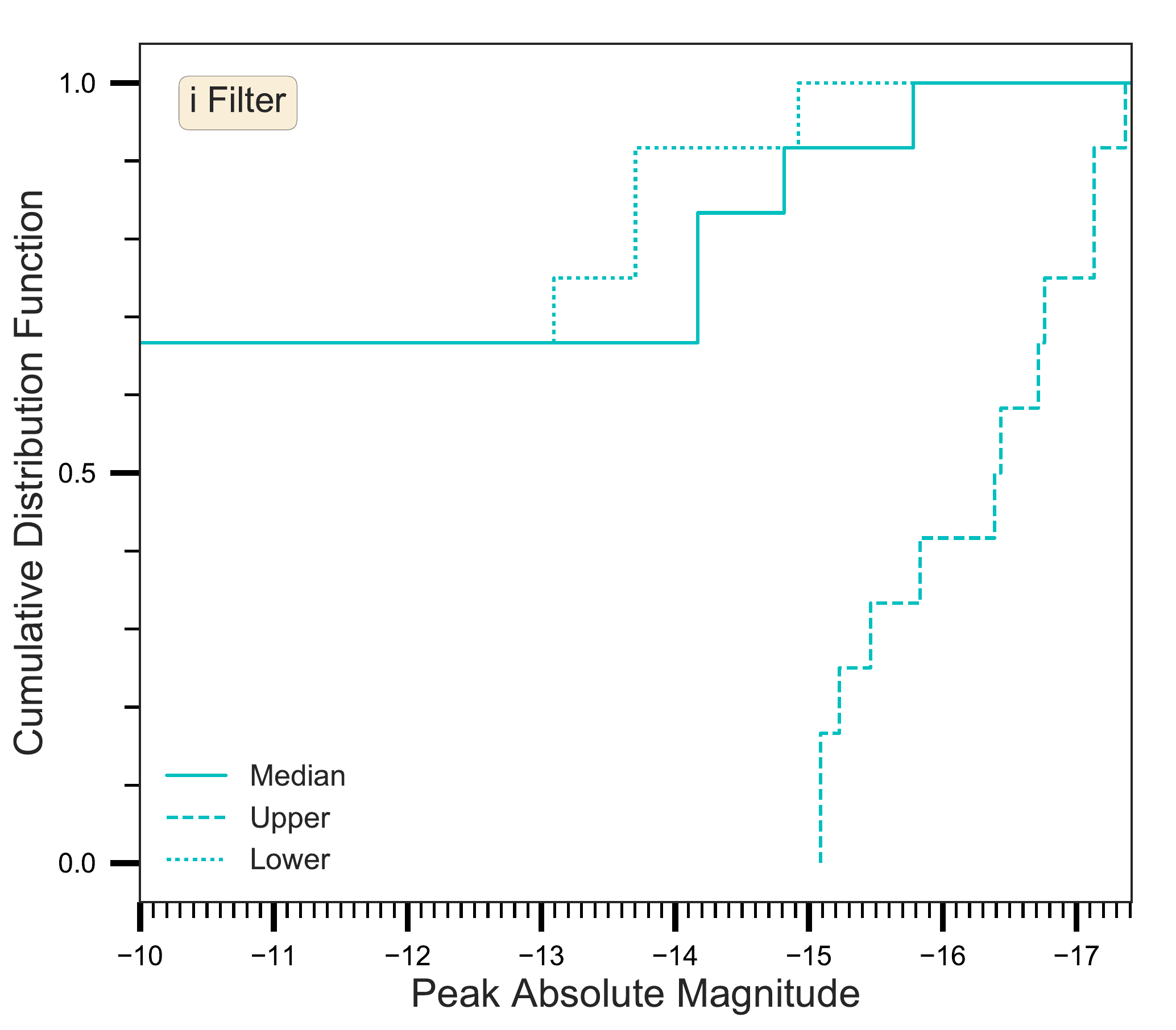}
    \includegraphics[width=3.3in]{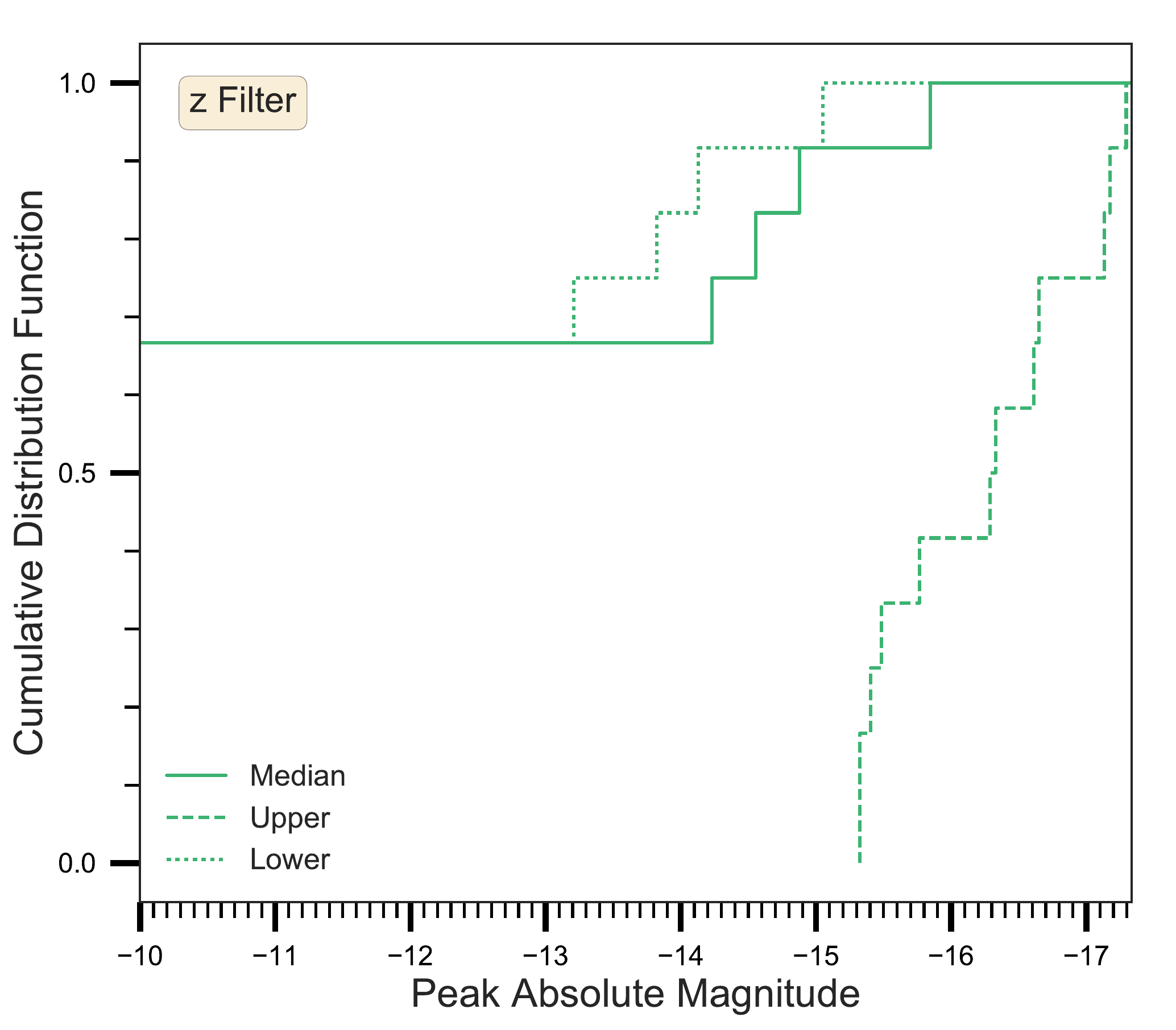}
    \includegraphics[width=3.3in]{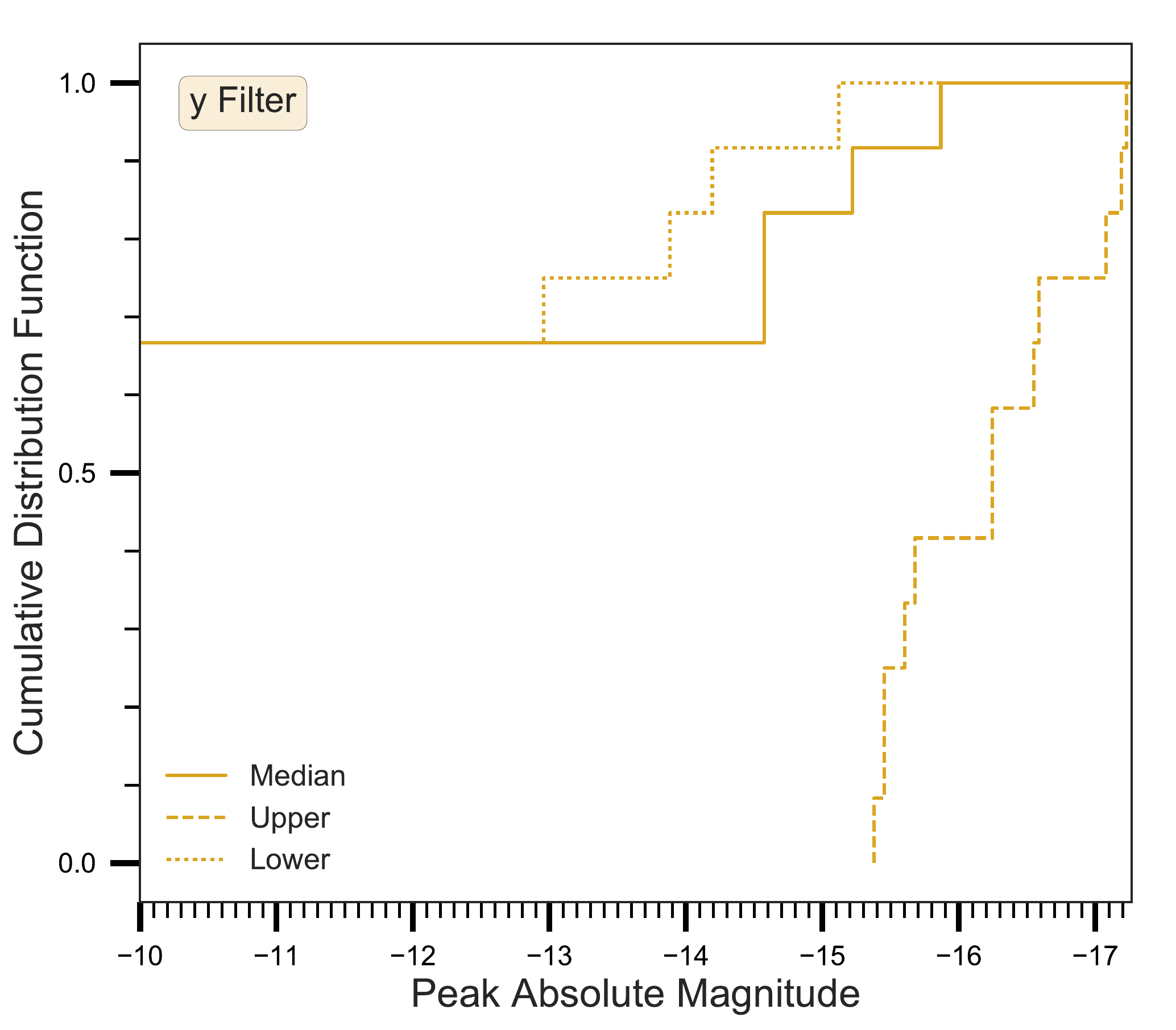}
    \includegraphics[width=3.3in]{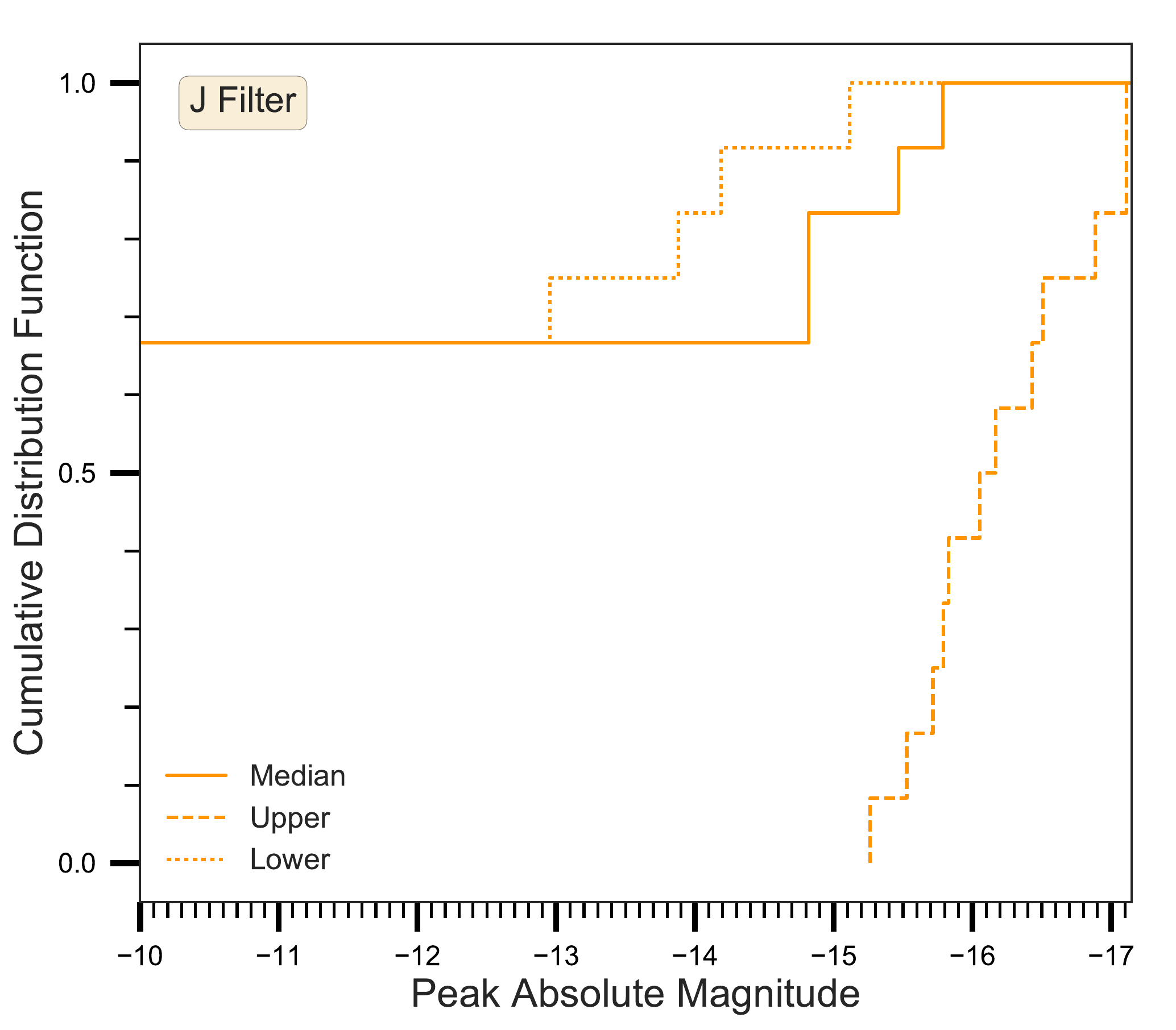}
\end{figure*}
\begin{figure*}[t]
	\centering
    \includegraphics[width=3.3in]{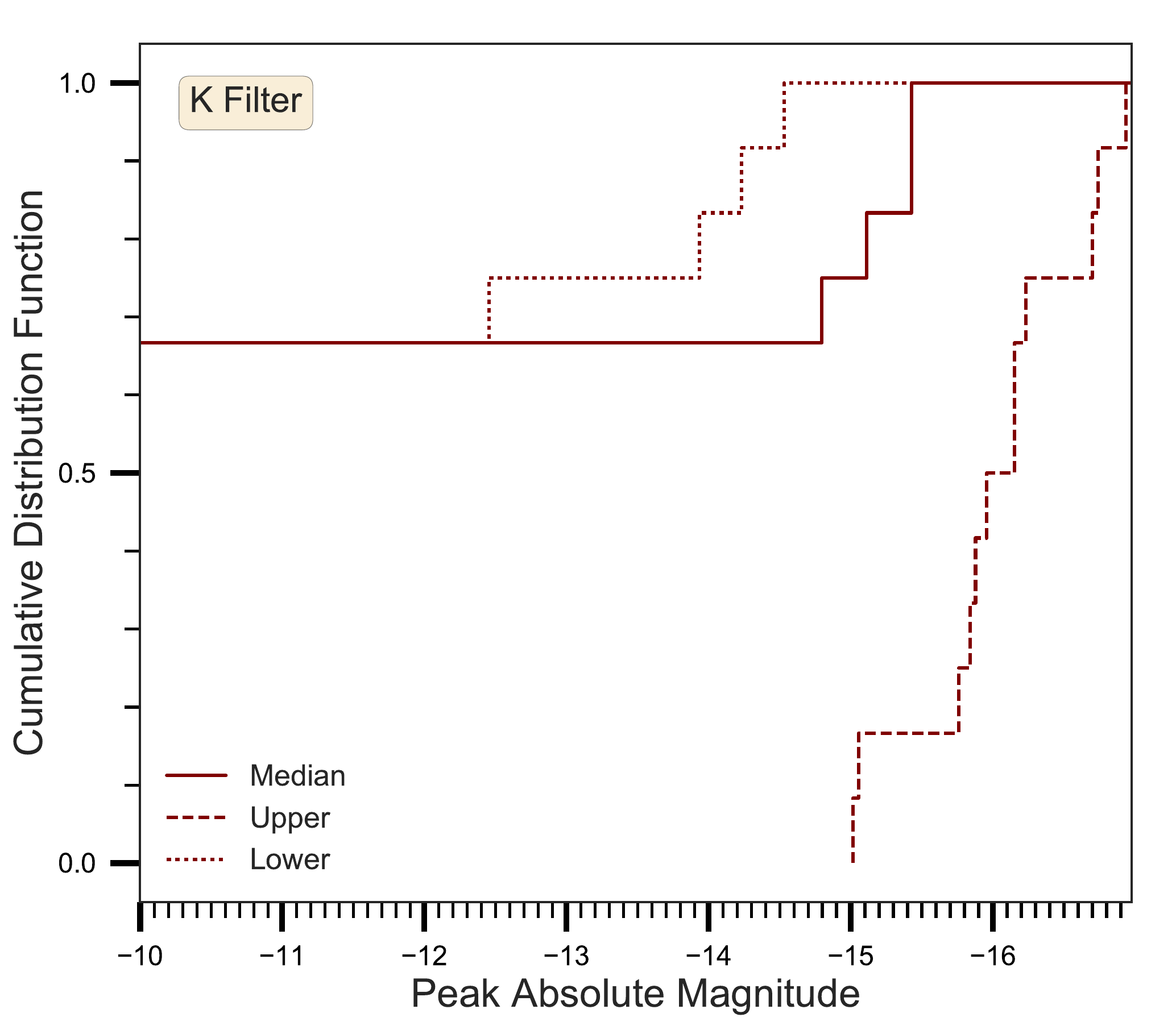}
    \caption{Luminosity distributions in u, r, i, z, y, J and K filters.}
    \label{fig:lum_func_allbands}
\end{figure*}

\begin{figure*}[t]
	\includegraphics[width=3.3in]{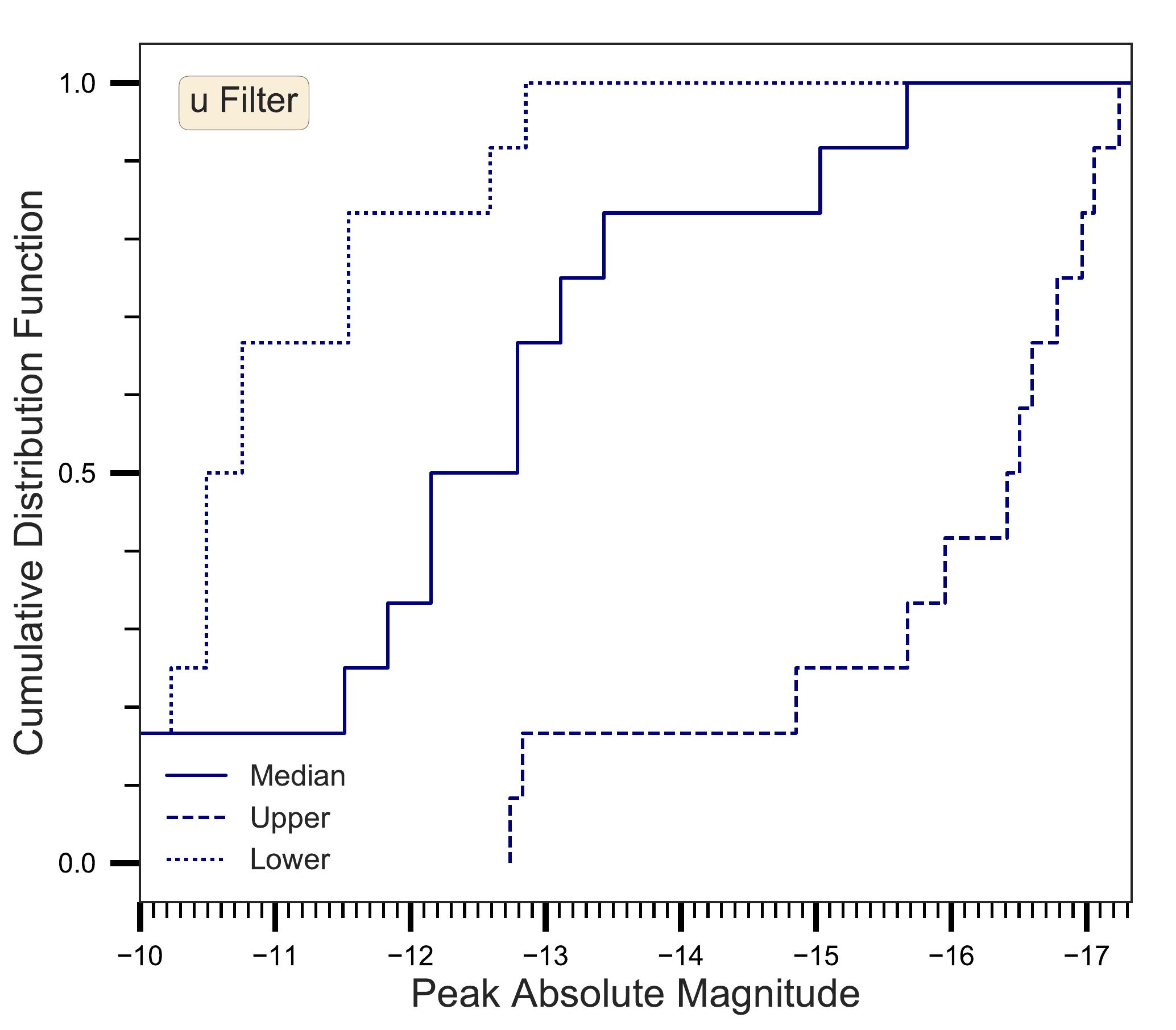}
    \includegraphics[width=3.3in]{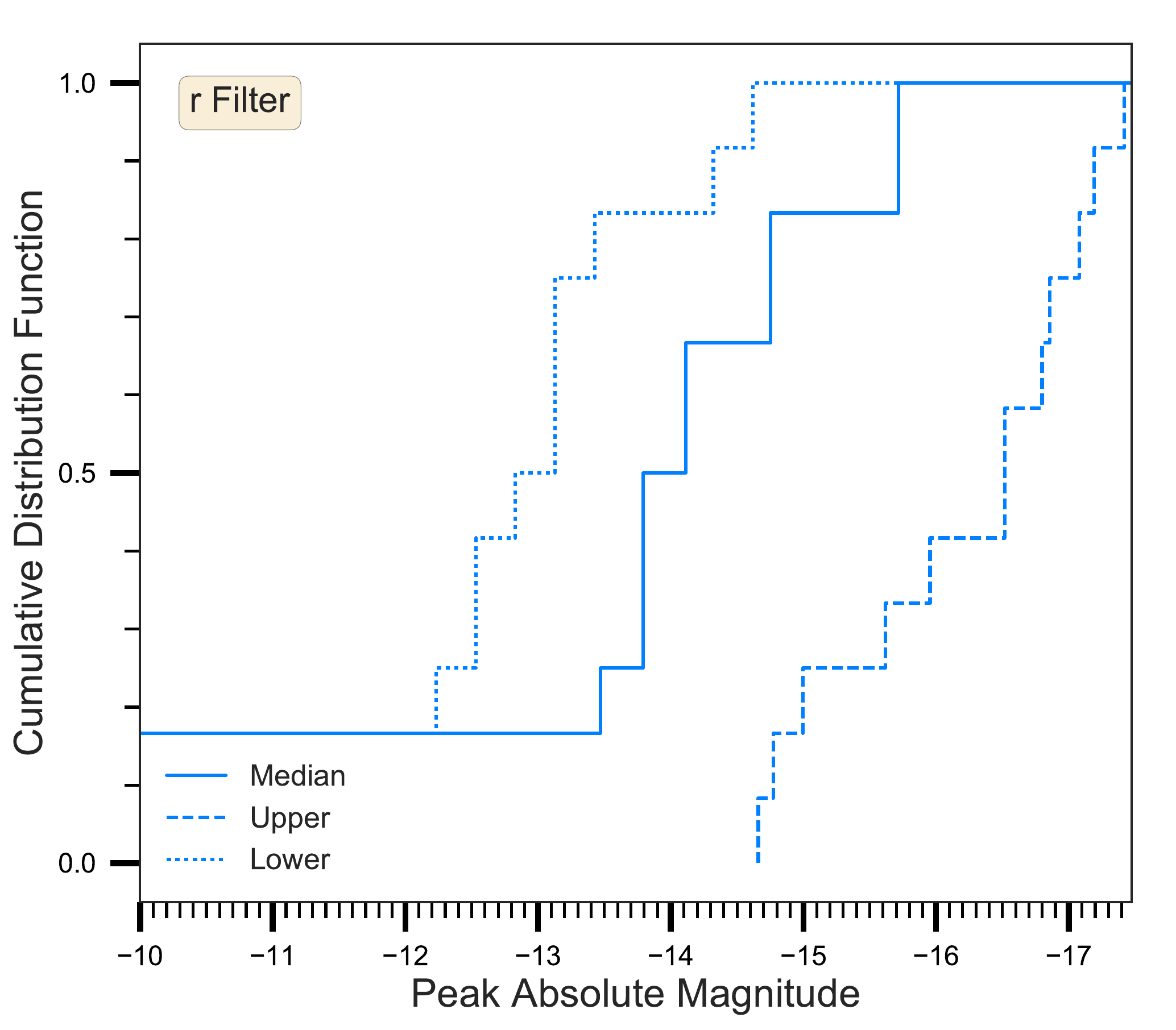}
    \includegraphics[width=3.3in]{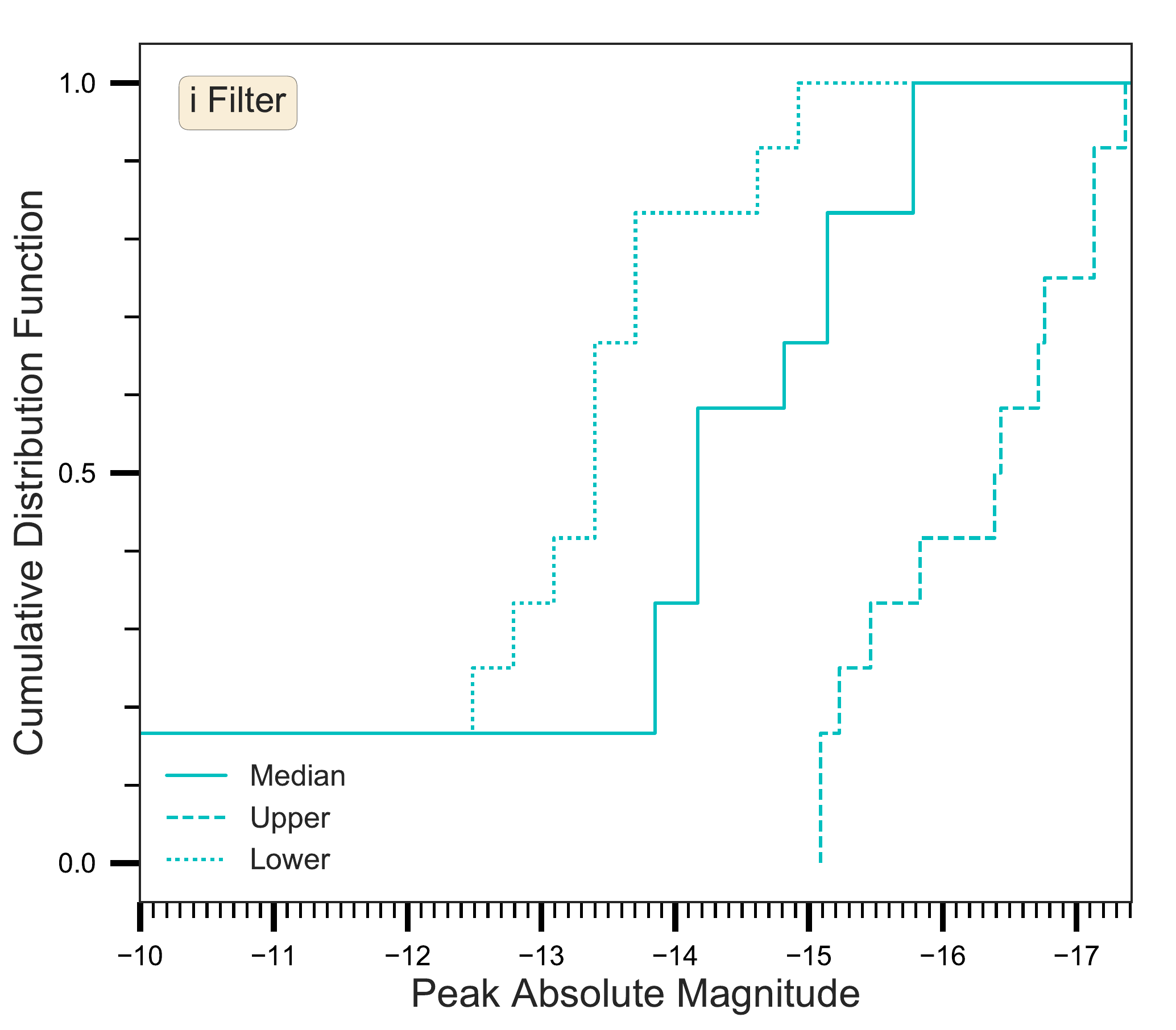}
    \includegraphics[width=3.3in]{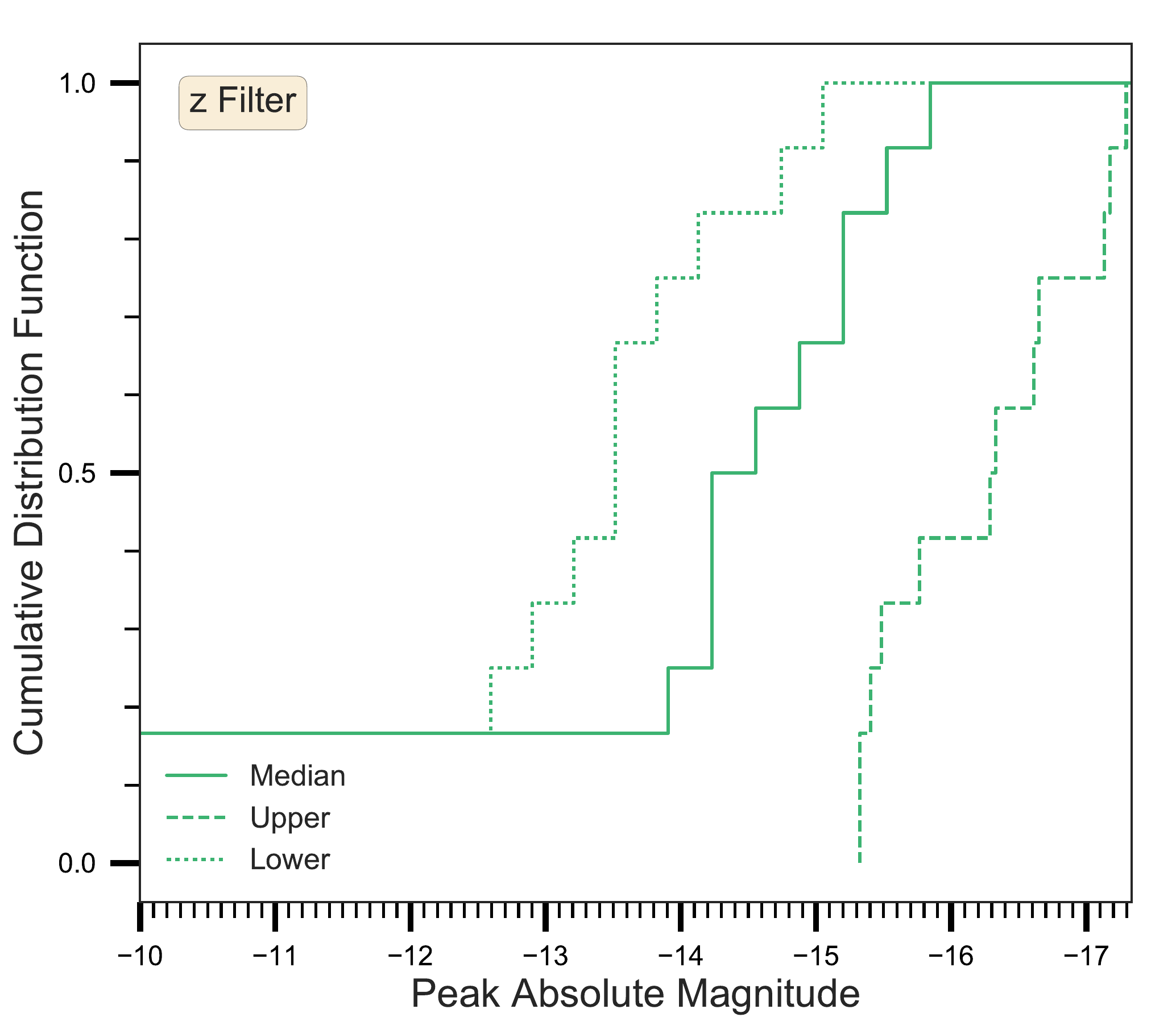}
    \includegraphics[width=3.3in]{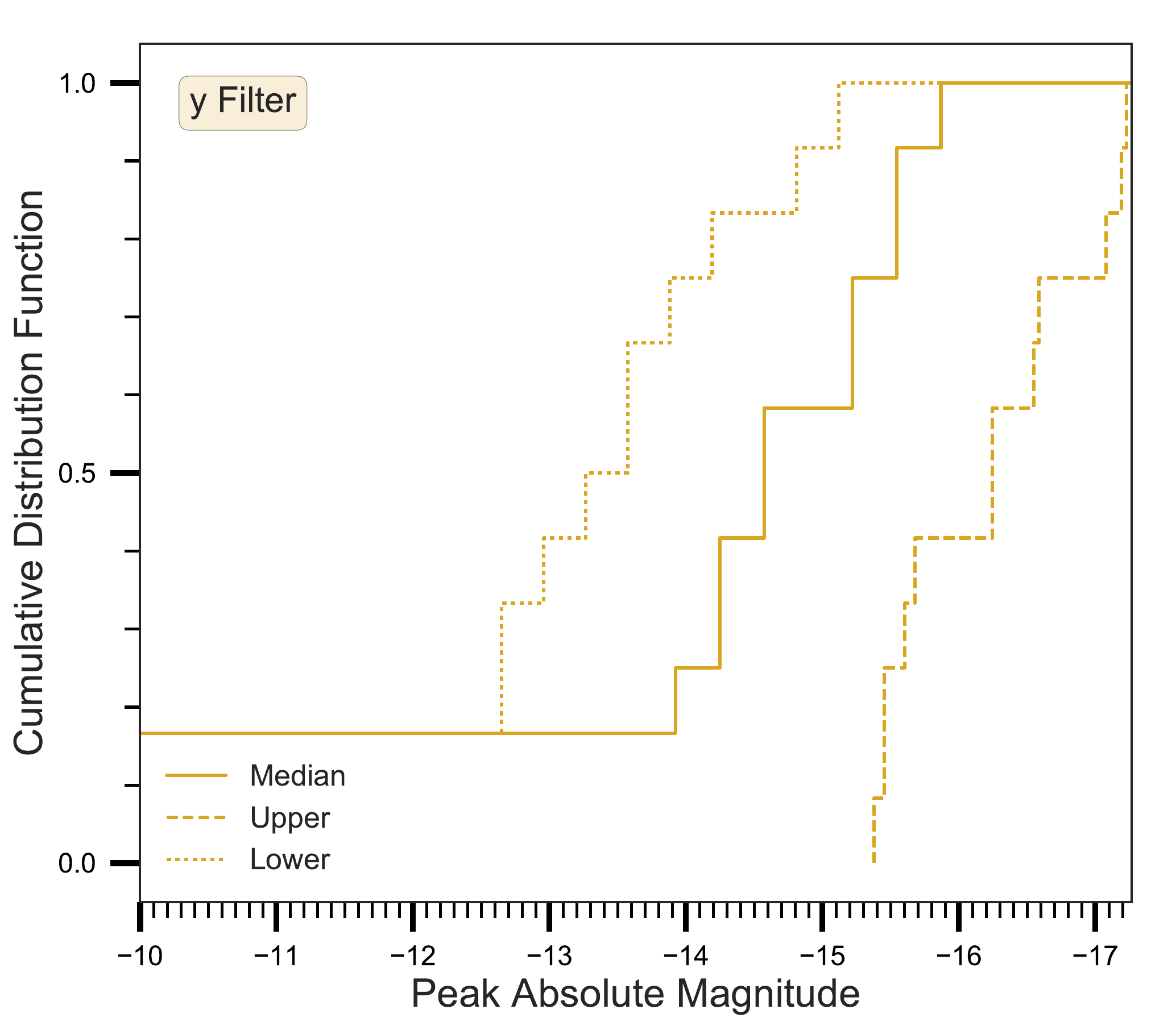}
    \includegraphics[width=3.3in]{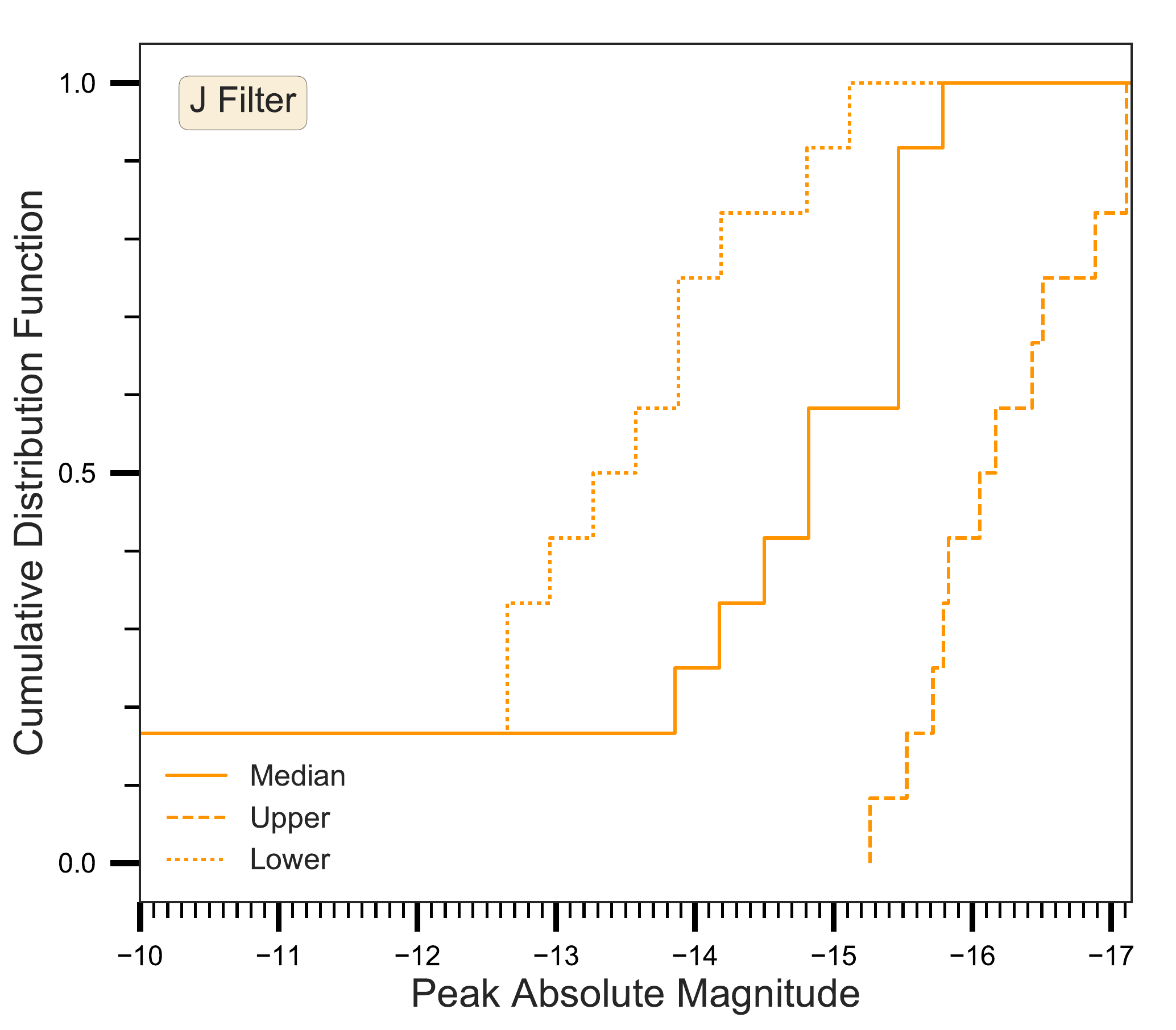}
\end{figure*}
\begin{figure*}[t]
     \centering
    \includegraphics[width=3.3in]{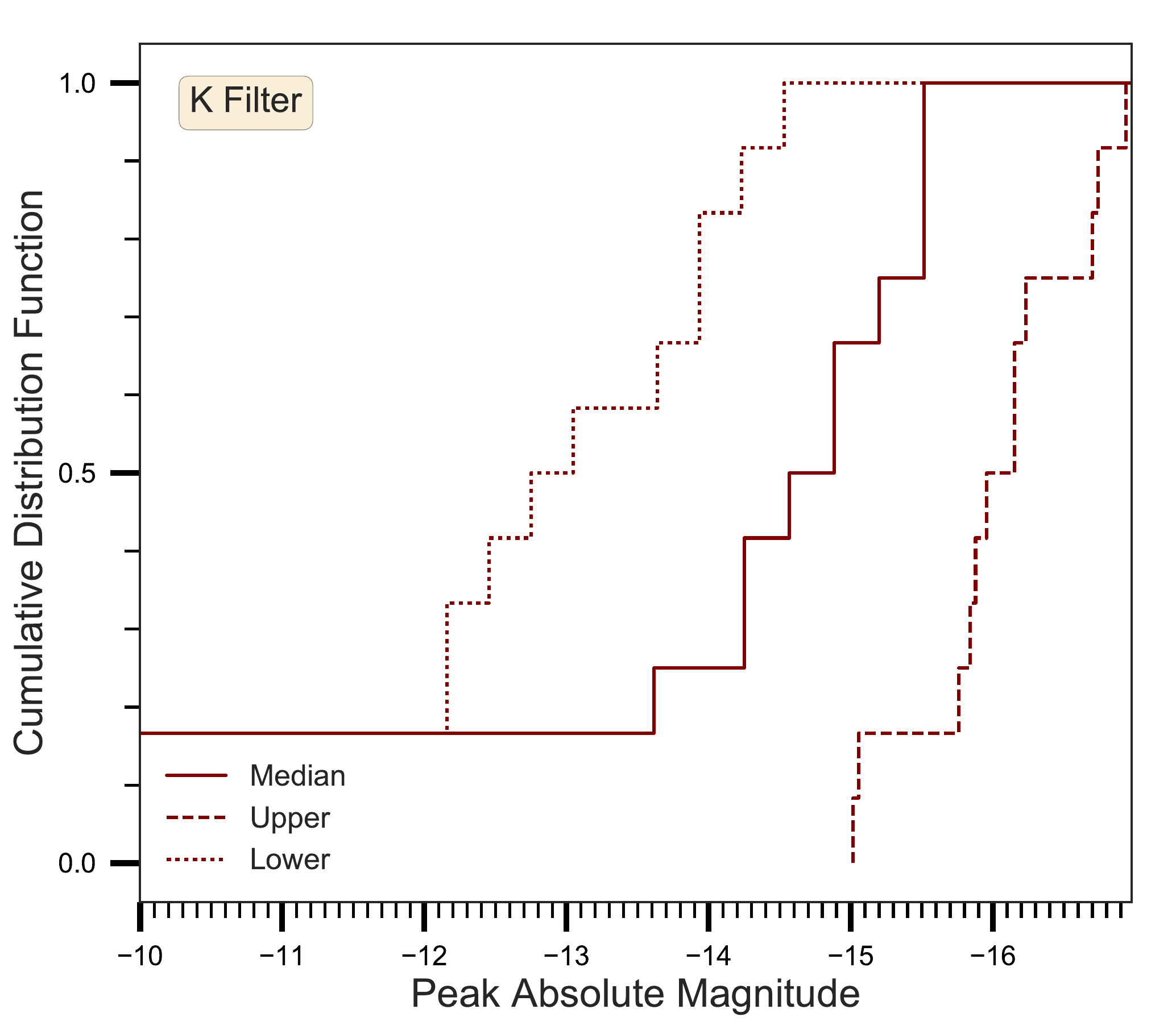}
    \caption{Same as Fig. \ref{fig:lum_func_allbands} but here we promoted GRB150101B, GRB050724A, GRB061201, GRB080905A, GRB150424A, GRB160821B to kilonova events.}
    \label{fig:lum_func_allbands_plus}
\end{figure*}

\end{document}